\documentclass[12pt]{article}
\usepackage{amsmath,amssymb,array,calc,epsfig}
\usepackage{graphicx}
\usepackage[all,cmtip]{xy}
\usepackage{cite}

\renewcommand{\tilde}{\widetilde}

\DeclareFontFamily{U}{rsf}{}
\DeclareFontShape{U}{rsf}{m}{n}{
<5> <6> rsfs5 <7> <8> <9> rsfs7 <10-> rsfs10}{}
\DeclareMathAlphabet\Scr{U}{rsf}{m}{n}

\newcommand{\del}{\partial}
\newcommand{\delbar}{{\bar\del}}

\newcommand{\bC}{{\mathbb{C}}}

\newcommand{\bM}{{\mathbb{M}}}
\newcommand{\bR}{{\mathbb{R}}}
\newcommand{\bP}{{\mathbb{P}}}

\newcommand{\bS}{{\mathbb{S}}}
\newcommand{\bT}{{\mathbb{T}}}

\newcommand{\cA}{{\mathcal{A}}}
\newcommand{\cB}{{\mathcal{B}}}

\newcommand{\cF}{{\mathcal{F}}}
\newcommand{\cG}{{\mathcal{G}}}
\newcommand{\cH}{{\mathcal{H}}}

\newcommand{\cL}{{\mathcal{L}}}
\newcommand{\cM}{{\mathcal{M}}}
\newcommand{\cN}{{\mathcal{N}}}
\newcommand{\cO}{{\mathcal{O}}}
\newcommand{\cP}{{\mathcal{P}}}
\newcommand{\cR}{{\mathcal{R}}}
\newcommand{\cS}{{\mathcal{S}}}
\newcommand{\cT}{{\mathcal{T}}}
\newcommand{\cY}{{\mathcal{Y}}}
\newcommand{\cV}{{\mathcal{V}}}

\newcommand{\scri}{{\Scr{I}}}

\newcommand{\e}{{\rm e}}
\newcommand{\im}{{\rm i}}

\newcommand{\rd}{{\mathrm d}}

\newcommand{\be}{\begin{equation}}
\newcommand{\ee}{\end{equation}}

\begin{document}

\title{Gravity, Twistors and the MHV Formalism}

\author{Lionel Mason \& David Skinner\\
\small{\it{The Mathematical Institute,}}\\
\small{\it{24-29 St.~Giles', Oxford OX1 3LP,}}\\
\small{\it{United Kingdom.}}}

\maketitle

\begin{abstract}
We give a self-contained proof of the formula for the MHV
amplitudes for gravity conjectured by Berends, Giele \& Kuijf and
use the associated twistor generating function to define a twistor
  action for the MHV diagram approach to gravity.

Starting from a background field calculation on a spacetime with anti self-dual
curvature, we obtain a simple spacetime formula for the scattering
of a single, positive helicity linearized graviton into one of
negative helicity.  Re-expressing our integral in terms of twistor
 data allows us to consider a spacetime that is asymptotic to a
 superposition of plane waves.  Expanding these out perturbatively
 yields the gravitational MHV amplitudes of Berends, Giele \& Kuijf.

We go on to take the twistor generating function off-shell at the
perturbative level. Combining this with a twistor action for the
anti self-dual background, the generating function provides the MHV
vertices for the MHV diagram approach to perturbative gravity. We
finish by extending these results to supergravity, in particular
$\cN=4$ and $\cN=8$.  

\end{abstract}

\section{Introduction}
\label{sec:intro}

Recent advances in understanding the perturbative structure of
gravity~(see {\it e.g.}
\cite{BDPR,BB-BD,BCDJKR,BCFIJ,BBST,Freddy,Benincasa,CachazoSkinner,FreddyNima,PierreEmil1,PierreEmil2, 
BBDIPR,BBDIPR2,Nasti,EF,BEF}) have uncovered structures that are not
visible in the standard spacetime formulation of general relativity.
A particularly striking development has been the chiral MHV (Maximal
Helicity Violating) diagram
formulation~\cite{BB-BD,BBDIPR,Nasti,BEF}.  In this approach, the
full perturbation theory for gravity, at least at tree level, is built
up out of standard massless scalar propagators and {\it MHV vertices}.
These vertices are off-shell continuations of amplitudes describing
interactions of $n$ linearized gravitons in momentum eigenstates, two
of which have positive\footnote{We will use Penrose conventions for
twistor space, in which the amplitudes supported on a twistor line
are `mostly minus'; these amplitudes are usually thought of as
$\overline{\rm MHV}$, but will be called MHV here. Our conventions
are detailed at the end of the introduction.} helicity while $n-2$
have negative helicity. Such amplitudes were first conjectured for 
Yang-Mills by Parke \& Taylor~\cite{PT} (and proved by Berends and
Giele~\cite{BG}) and later a more complicated formula \eqref{BGK} for gravity was
conjectured by Berends, Giele \&
Kuijf~\cite{BGK}.

\begin{figure}
\begin{center}
\includegraphics[height=55mm]{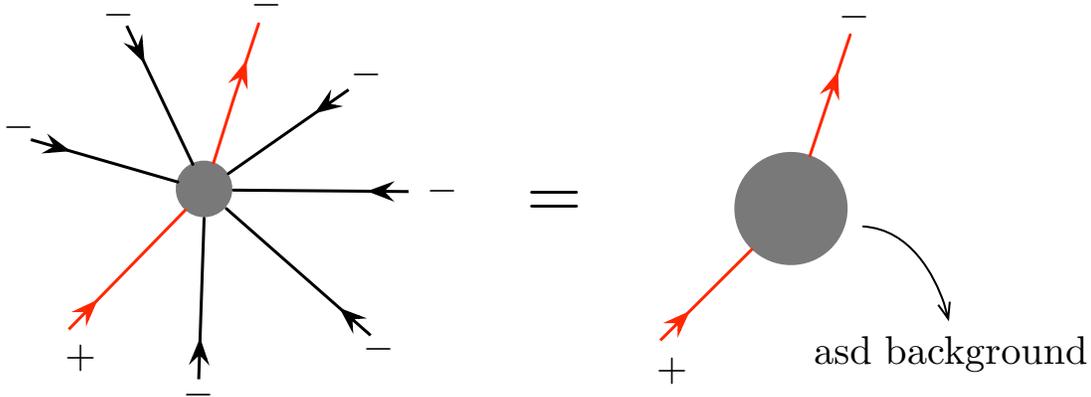}
\caption{{\it Reversing the momentum of one of the positive helicity
particles leads to the interpretation of the MHV amplitude as
measuring the helicity-flip of a single particle which traverses a
region of ASD background curvature.}}
\label{fig:background}
\end{center}
\end{figure}

Both in gravity and Yang-Mills, MHV amplitudes are considerably
simpler than a generic tree-level helicity amplitude. In particular,
they may involve an arbitrary number of negative helicity gravitons
(gluons) at little or no cost in complexity. Why should this be?
Bearing in mind that a negative helicity graviton that has positive
frequency is anti self-dual~\cite{birula,ashhelicity}, the 
picture in figure~\ref{fig:background} interprets MHV
amplitudes as measuring the helicity-flip of a single particle as it
traverses a region of anti self-dual (ASD) background curvature. The
asd Einstein equations, like the ASD Yang-Mills equations, have long
been known to be completely integrable~\cite{Penrose,HNPT} and lead to
trivial scattering at tree-level.  From this perspective, the key
simplification of the MHV formalism arises because the ASD background,
despite its non-linearities, can effectively be treated as a free
theory. The MHV amplitudes themselves represent the first departure
from anti self-duality.

The MHV formulation is essentially chiral. For gravity, this chirality
suggests deep links to Plebanski's chiral
action~\cite{Plebanski,CDM,MasonFrauen}, to Ashtekar
variables~\cite{MasonFrauen,Ashtekar} and to twistor
theory~\cite{Penrose,PenroseMaccallum}. It is the purpose of this
article to elucidate these connections further and to go some way
towards a non-linear formulation that helps illuminate the underlying
nonperturbative structure.  Thus we begin in section~\ref{sec:gravity}
with a brief review of the Plebanski action, explaining how it can be
used to expand gravity about its anti self-dual sector. Similar
discussions have been given in~\cite{Plebanski,AJS} and more
recently~\cite{AH} whose treatment we follow most closely.

On an ASD background, a linearized graviton has a canonically defined
self-dual part, but its anti self-dual part shifts as it moves through
the spacetime. We show in section~\ref{sec:asd} that the tree-level
amplitude for this shift to occur is precisely measured by a simple
spacetime integral formula.  This integral is a generating function
for all the MHV amplitudes. To obtain them in their usual form, one
must expand out the background field in terms of fluctuations around
flat spacetime. Understanding how a non-linear anti self-dual field is
composed of linearized gravitons is feasible precisely because the asd
equations are integrable, but nonetheless the inherent non-linearity
makes this a rather complicated task on spacetime~\cite{Rosly}. However, by going
to twistor space and using Penrose's non-linear graviton
construction~\cite{Penrose}, the ASD background can be reformulated in
an essentially linear way. Hence in section~\ref{sec:twistors}, after
reviewing the relevant twistor theory of both linear gravity and
non-linear ASD gravity, we obtain a twistor representation of the
generating function using twistor integral formul\ae\ for the
spacetime fields. We will see that it is straightforward to construct
a twistor space for a non-linear ASD spacetime that asymptotically is
a linear superposition of momentum eigenstates.  This uses a
representation for the twistor space as an asymptotic twistor space
constructed from the asymptotic data and is closely related to
Newman's $\Scr{H}$-space construction~\cite{Newman}. Thus, we can use the
twistor description to expand our generating function around Minkowski
spacetime.  A completely analogous story is true in
Yang-Mills~\cite{Mason,BMS}, with the corresponding twistor expression
yielding all the Parke-Taylor amplitudes. This is reviewed in
appendix~\ref{sec:ym}; some readers may find it helpful to refer to
the (somewhat simpler) Yang-Mills case for orientation.

Performing the expansion, one finds that the $n$-point amplitude comes
from an integral over the space of holomorphic twistor lines with $n$
marked points. The marked points support operators representing the
external gravitons; the $2$ positive helicity gravitons are
represented by 1-form insertions while the $n-2$ negative helicity
gravitons give insertions of vector fields. These vectors
differentiate the external wavefunctions, leading to what is sometimes
called `derivative of a $\delta$-function support'. The 1-forms and
vector fields really represent elements of certain cohomology classes
on twistor space. It is interesting to note that these are the same
cohomology groups that arise as (part of) the BRST cohomology in
twistor-string theory~\cite{BW,het}, but here there are extra
constraints which ensure that they represent Einstein, rather than
conformal, gravitons.  A string theory whose vertex operators satisfy
these extra constraints was constructed in~\cite{AHM}, although these
models do not appear to reproduce the MHV amplitudes~\cite{Nair2}.

Integrating out the twistor variables finally yields the formula
\begin{multline}
\cM^{(n)}=\frac{\kappa^{n-2}}{\hbar}\,\delta^{(4)}\left(\sum_{i=1}^n p_i\right)\\
\ \times \left\{\frac{[1n]^8}{[1\,n-1][n-1\,n][n\,1]}\frac{1}{C(n)}
  \prod_{k=2}^{n-1}\frac{\langle k|p_{k+1}+\cdots +p_{n-1}|n]}{[kn]}
  + {\rm P}_{\{2,\ldots,n-2\}}\right\}\ ,
\label{amplitudes}
\end{multline}
for the $n$-particle amplitude $\cM^{(n)}$, where
$\kappa=\sqrt{16\pi{\rm G_N}}$ and we have used the spinor-helicity
formalism: the $i^{\rm th}$ external graviton is taken to have null
momentum $p_{\alpha\dot\alpha}^{(i)}=|i\rangle[i|$, where $|i\rangle$
and $[i|$ respectively denote the anti-self-dual and self-dual spinor
constituents of $p_i$, and $C(n)$ is the cyclic product
$[12][23]\cdots[n-1\,n][n1]$. The symbol ${\rm P}_{\{2,\ldots,n-2\}}$
denotes a sum over permutations of gravitons $2$ to $n-2$; the
amplitude is completely symmetric in the external states (up to the
overall factor $[1n]^8$ from the two positive helicity gravitons) once
these permutations are accounted for.  Equation~(\ref{amplitudes}) is
not the original expression of BGK~\cite{BGK} and an analytic proof
that the two forms coincide for arbitrary $n\geq4$ is given in
appendix~\ref{sec:BGK}. The twistor formula also yields the correct
3-point amplitude, which is non-zero in complexified momentum space
(although yields zero on a Lorentzian real slice). Our generating
function may be simply extended to the case of MHV amplitudes in
supergravity, and this is discussed in section~\ref{sec:sugra} for
$\cN=4$ and $\cN=8$ supergravity.

In the {\it MHV diagram formalism}, the full perturbation theory is
reproduced from MHV amplitudes that are continued off-shell to provide
vertices. These vertices are then connected together with propagators joining
positive and negative helicity lines. With $p$ such propagators, one
obtains a ${\rm N}^p{\rm MHV}$ amplitude, usually thought of in terms
of the scattering of $2+p$ positive helicity gravitons and an
arbitrary number of negative helicity gravitons. In
section~\ref{sec:twistact} we continue our twistorial generating
function off-shell and couple it to the twistor action for anti
self-dual gravity constructed in~\cite{MW}. The Feynman diagrams of
the resulting action reproduce (in a certain gauge) the MHV diagram
formalism for gravity. At present, we understand this action only in
perturbation theory, and its validity as an action for gravity rests on the validity of the MHV diagram formalism. 
It would be very interesting to learn how the off-shell twistor action generates
off-shell curved spacetime metrics, or to see if the existence of the twistor action implies that the MHV expansion is indeed valid.

\smallskip

The gravitational MHV amplitudes were originally calculated~\cite{BGK} using
the Kawai, Llewelyn \& Tye relations~\cite{KLT}, and subsequently recalculated in a different form using
the Britto, Cachazo, Feng \& Witten recursion relations~\cite{BCFW},
suitably modified for gravity~\cite{BBST,Freddy,Benincasa}.  Although the BGK expression is
strongly constrained by having the correct soft and collinear limits, strictly speaking, BGK
were only able to prove that their formula followed from the KLT relations for $n\leq11$ external
particles. The formul\ae~obtained from BCFW recursion relations have
also only been verified to be equivalent to the BGK expression up to this level.
Our derivation is a complete constructive proof of
the BGK formula (the formul\ae~of~\cite{BBST,Freddy,Benincasa} are also independently
proved).  Evidence for a MHV diagram formulation of perturbative
gravity has been discussed in~\cite{BBDIPR,Nasti,BEF}, based on
recursion relations. It has been established~\cite{BEF} that the MHV
diagrams yield the correct $n$-graviton amplitudes, again for $n\leq11$. Reference~\cite{BEF} also gives a generating function for MHV amplitudes in $\cN=8$ supergravity, taking the BGK amplitudes as an
input.

Some steps towards an MHV action for gravity have been taken in~\cite{ATh}, starting from lightcone gauge in spacetime and inspired by the work of Mansfield in Yang-Mills~\cite{Mansfield,Ettle}.  A twistorial generating function which
reproduces the gravity MHV amplitudes was constructed by Nair in~\cite{Nair3}. Nair's paper has influenced this one; the main difference is that we give an independent derivation of the amplitudes, starting from a spacetime formula for scattering off an ASD background. We also take a more geometrical
perspective than~\cite{Nair3}. A treatment of the MHV amplitudes that
emphasizes their close connection to the integrability of asd
backgrounds has been given in~\cite{Rosly,SelivanovYM} using
`perturbiners'.

\subsection{Conventions and notation}
\label{conventions}

Flat Minkowski spacetime $\bM$ is taken to be $\bR^4$ with metric of
Lorentz signature $(+---)$ and with vector indices $a=0,1,2,3$.  Let
$\mathbb{S}^+$ and $\mathbb{S}^-$ be the self-dual and anti self-dual
spin spaces. Elements of $\mathbb{S}^\pm$ will be taken to have dotted
and undotted Greek indices respectively, {\it i.e.} $\dot\alpha, \ldots=\dot 0,
\dot 1$; $\alpha, \ldots=0,1$.  We denote the Levi-Civita alternating
spinor by $\varepsilon_{\alpha\beta}=\varepsilon_{[\alpha\beta]}$, with
$\varepsilon_{01}=-1$, {\it etc.}  We often use the notation
$r^\alpha\leftrightarrow|r\rangle$ and
$s^{\dot\alpha}\leftrightarrow|s]$ and then $[p\,r]=p^\alpha
r^\beta\varepsilon_{\alpha\beta}$ and $\langle
s\,t\rangle=s^{\dot\alpha}t^{\dot\beta}\varepsilon_{\dot\alpha\dot\beta}$
denote the $SL(2,\bC)$-invariant inner products.  In complexified
spacetime the two spin bundles will also be denoted $\mathbb{S}^+$ and
$\mathbb{S}^-$.  On a Lorentzian real slice they are related by
complex conjugation $\overline{\mathbb{S}^+}=\mathbb{S}^-$, which
therefore 
exchanges dotted and undotted spinor indices.  Vector indices
$a=0,1,2,3$ can be replaced by spinor indices, so that the position
vector of a point can be given as 
\be
x^{\alpha\dot\alpha}= \frac1{\sqrt2}\begin{pmatrix} x^0+x^3&x^1+ix^2\\
					 x^1-ix^2&x^0-x^3\end{pmatrix} \ .
\label{x}\ee
The Lorentz reality condition is
$x^{\alpha\dot\alpha}=\bar x^{\dot\alpha\alpha}$, so that the rhs of~\eqref{x} is a Hermitian matrix.  We will often
work on complexified spacetime, where $x^a$ and $x^{\alpha\dot\alpha}$ are
complex and the reality condition is dropped.

Projective twistor space $\bP\bT^\prime$ is the space of totally
null self-dual two planes ($\alpha$-planes) in complexified
spacetime. We describe $\bP\bT^\prime$ using homogeneous coordinates
$(\omega^{\alpha},\pi_{\dot\alpha})$, with the incidence relation
being $\omega^\alpha=\im x^{\alpha\dot\alpha}\pi_{\dot\alpha}$; the
solutions for $x^{\alpha\dot\alpha}$ holding
$(\omega^\alpha,\pi_{\dot\alpha})$ constant defines the
$\alpha$-plane. In these conventions, an element of
$H^1(\mathbb{PT}^\prime, \cO(-2s-2))$ corresponds to an on-shell
massless field of helicity $s$ in spacetime by the Penrose
transform. Thus a {\it negative} helicity gluon has homogeneity zero
in twistor space, and the amplitudes supported on degree 1 holomorphic
curve are `mostly minus'. We call such $\langle++--\cdots--\rangle$
amplitudes MHV, although they are the complex conjugate of what is
called an MHV amplitude in much of the scattering theory
literature. With our conventions, Witten's twistor-string
theory~\cite{Witten} is really in dual twistor space.  In Lorentzian
signature, twistor space and its dual are related via complex
conjugation, {\it i.e.}
$(\omega^\alpha,\pi_{\dot\alpha})\in\bP\bT^\prime\mapsto(\bar\pi_{\alpha},\bar\omega^{\dot\alpha})\in{\bP\bT^\prime}^*$,
reflecting the Lorentzian conjugation of Weyl spinors. For
complexified spacetime, one often gives dual twistor space independent
coordinates $(\lambda_\alpha,\mu^{\dot\alpha})$ which are the
coordinates used in~\cite{Witten}.


\section{MHV Amplitudes on ASD Background Fields}
\label{sec:gravity}

\subsection{The Plebanski action}
The (complexified) spin group of a Lorentzian four manifold $M$ is
$SL(2,\bC)\times SL(2,\bC)$. Correspondingly, the tangent bundle $TM$
decomposes into the self-dual and anti self-dual spin bundles
$\mathbb{S}^\pm$ as $TM\simeq\mathbb{S}^+\otimes\mathbb{S}^-$. Each
$SL(2,\bC)$ factor acts non-trivially on only either $\mathbb{S}^+$ or
$\mathbb{S}^-$ and so any connection on $TM$ may be decomposed into
connections on the two spin bundles as $
\Gamma\oplus\widetilde\Gamma$. Splitting the curvature two-form into
its self-dual and anti self-dual parts $R^\pm$, one finds that
$R^+=R^+(\Gamma)$ and $R^-=R^-(\widetilde\Gamma)$ so that the
self-dual (ASD) part of the curvature depends only on the connection
on $\mathbb{S}^+$ $(\mathbb{S}^-)$. (On a Lorentzian four-manifold
$*^2=-1$, so the SD/ASD curvatures are complex and
$\overline\Gamma=\widetilde{\Gamma}$, $R^+=\overline{R^-}$. In
Euclidean or split signature the spin connections and $R^\pm$ are real
and independent. We will mostly work on complexified spacetime,
imposing reality conditions only at the end.)

Plebanski~\cite{Plebanski} gave a chiral action for Einstein's general
relativity that brings out this structure (see
also~\cite{CDM,MasonFrauen}).  In his approach, the basic variables
are the self-dual spin connection $\Gamma$, together with a tetrad of
1-forms $e^{\alpha\dot\alpha}$ which define the metric by
\begin{equation}
ds^2=\varepsilon_{\alpha\beta}\varepsilon_{\dot\alpha\dot\beta}\,e^{\alpha\dot\alpha}\,e^{\beta\dot\beta}\ ,
\label{metricdef}
\end{equation}
where $\varepsilon_{\alpha\beta}=\varepsilon_{[\alpha\beta]}$,
$\varepsilon_{01}=1$ and similarly for  
$\varepsilon_{\dot\alpha\dot\beta}$. The components of the tetrad are
defined by $e^{\alpha\dot\alpha} = e^{\alpha\dot\alpha}_a dx^a$ and
form a vierbein.  Plebanski's action is a first-order theory in which
$\Gamma$ and the tetrad are treated as independent {\it a
priori}. In the absence of a cosmological constant, the action is  
\begin{equation}
S[\Sigma,\Gamma ]=\frac{1}{\kappa^2}\int_M \Sigma^{\dot\alpha\dot\beta}\wedge
\left(\rd\Gamma + \Gamma\wedge\Gamma\right)_{\dot\alpha\dot\beta}
\label{gravityact}
\end{equation}
where $\kappa^2=16\pi{\rm G_N}$ and $\Sigma^{\dot\alpha\dot\beta}$ are
three self-dual two-forms, given in terms of the tetrad by
$\Sigma^{\dot\alpha\dot\beta}=e^{\alpha(\dot\alpha}_{\phantom\beta}\wedge
e_\alpha^{\ \dot\beta)}$.  It is a striking fact that
$\widetilde\Gamma$ plays no role in this action\footnote{Of
course, one can still construct an ASD spin connection from the
tetrad.}. It nevertheless describes full (non-chiral) Einstein
gravity, as follows from the field equations
\begin{eqnarray}
\rd\Sigma^{\dot\alpha\dot\beta}+2
\Gamma^{(\dot\alpha}_{\ \,\dot\gamma}\wedge\Sigma^{\dot\beta)\dot\gamma}_{\phantom{()}}&=&0
\label{graveom1}\\
\left(\rd\Gamma_{\dot\alpha\dot\beta}+
\Gamma^{\dot\gamma}_{\ (\dot\alpha}\wedge\Gamma^{\phantom{\dot\gamma}}_{\dot\beta)\dot\gamma}\right)
	\wedge e^{\alpha\dot\alpha}&=&0\ .
\label{graveom2}
\end{eqnarray}
The first of these is the condition that $\Gamma$ is torsion-free,
which fixes it in terms of the tetrad. Since (after an integration by
parts) $\Gamma$ appears in the action only algebraically, this
equation may be viewed as a constraint. Imposing it
in~(\ref{graveom2}) implies that the Ricci curvature of the
metric~(\ref{metricdef}) vanishes, so that $M$ satisfies the vacuum
Einstein equations. Thus Plebanski's action is equivalent to the
Einstein-Hilbert action (upto a topological term).

It is also possible to take $\Sigma^{\dot\alpha\dot\beta}$ to be an
arbitrary set of self-dual 2-forms and view them as the basic
variables, as was done in~\cite{CDM}. The condition that
$\Sigma^{\dot\alpha\dot\beta}$ comes from a tetrad ({\it i.e.}
$\Sigma^{\dot\alpha\dot\beta}=e_{\phantom{\alpha}}^{\alpha(\dot\alpha}\wedge
e_\alpha^{\ \dot\beta)}$) is ensured by including a Lagrange
multiplier to enforce
$\Sigma^{(\dot\alpha\dot\beta}\wedge\Sigma^{\dot\gamma\dot\delta)}=0$. In
the present paper, this constraint will naturally be solved as part of
the construction of $\Sigma^{\dot\alpha\dot\beta}$ from twistor
space. We also remark that $\Sigma^{\dot\alpha\dot\beta}$ and
$\Gamma^{\dot\alpha}_{\ \dot\beta}$ may be thought of as a 4-covariant
form of Ashtekar variables~\cite{Ashtekar}: if $C$ is a spacelike
Cauchy surface in $M$, then the restriction of
$\Sigma^{\dot\alpha\dot\beta}$ to $C$ gives Ashtekar's densitized
triads\footnote{$i,j,k,\ldots$ are indices for the tangent space to
$C$.} $\sigma^{\dot\alpha\dot\beta\, i}_{[jkl]}$ via
\begin{equation}
\left.\Sigma^{\dot\alpha\dot\beta}_{[ij]}\right|_C =
3\sigma^{\dot\alpha\dot\beta\,k}_{[ijk]}\qquad 
\left.\Sigma^{\dot\alpha\dot\beta}_{[jk}\delta^i_{\ l]}\right|_C =
\sigma^{\dot\alpha\dot\beta\,i}_{[jkl]}\ , 
\label{Ashtekar}
\end{equation}
whereas the restriction of $\Gamma$ to $C$ is the
Ashtekar-Sen-Witten connection (see~\cite{MasonFrauen} for details).


\subsection{Linearizing around an anti self-dual background}
\label{sec:asd}

We will be particularly interested in {\it anti self-dual} solutions
to~\eqref{graveom1}-\eqref{graveom2}. On an ASD solution, the
self-dual spin bundle $\bS^+\to M$ is flat, so $\Gamma$ vanishes
upto a gauge transform. The torsion-free constraint~\eqref{graveom1}
becomes
\begin{equation}
		\rd\Sigma^{\dot\alpha\dot\beta}=0\ ,
\label{graveomasd}
\end{equation}
so that the {\it self-dual} part of the spin connection constructed from the tetrad $e^{\alpha\dot\alpha}$ must also be pure gauge. There are no constraints on the {\it anti} self-dual part of this connection, so the associated Riemann tensor 
$R^a_{\ bcd}(e)$ need not vanish, but is purely asd. Decomposing a general Riemann tensor into irreducibles 
gives~\cite{penrose-rindler}
\begin{equation}
\begin{aligned}
&R_{abcd} =
\Psi_{\alpha\beta\gamma\delta}\varepsilon_{\dot\alpha\dot\beta}
\varepsilon_{\dot\gamma\dot\delta}
+\tilde\Psi_{\dot\alpha\dot\beta\dot\gamma\dot\delta}\varepsilon_{\alpha\beta}
\varepsilon_{\gamma\delta}
+\Phi_{\alpha\beta\dot\gamma\dot\delta}\varepsilon_{\dot\alpha\dot\beta}
\varepsilon_{\gamma\delta}
+\Phi_{\gamma\delta\dot\alpha\dot\beta}\varepsilon_{\alpha\beta}
\varepsilon_{\dot\gamma\dot\delta}\\ 
&\hspace{1.35cm} + \frac{R}{12}\left(
\varepsilon_{\alpha\gamma}\varepsilon_{\beta\delta}
\varepsilon_{\dot\alpha\dot\beta}\varepsilon_{\dot\gamma\dot\delta}
+\varepsilon_{\alpha\beta}\varepsilon_{\gamma\delta}
\varepsilon_{\dot\alpha\dot\gamma}\varepsilon_{\dot\beta\dot\delta}\right)
\end{aligned}
\end{equation}
where $\tilde\Psi_{\dot\alpha\dot\beta\dot\gamma\dot\delta}=\tilde
\Psi_{(\dot\alpha\dot\beta\dot\gamma\dot\delta)}$ and
$\Phi_{\dot\alpha\dot\beta\gamma\delta}=\Phi_{(\dot\alpha\dot\beta)(\gamma\delta)}$
are the spinor forms of the self-dual part of the Weyl tensor and the
trace-free part of the Ricci tensor, respectively, and $R$ is the
scalar curvature.  With vanishing cosmological constant, $R^a_{\
bcd}(e)$ is anti self-dual if and only if
$\tilde\Psi_{\dot\alpha\dot\beta\dot\gamma\dot\delta}$,
$\Phi_{\dot\alpha\dot\beta\gamma\delta}$ and $R$ vanish. The ASD part
$\Psi_{\alpha\beta\gamma\delta}$ of the Weyl tensor need not vanish
(at least in complexified or Euclidean spacetime), but it obeys
$\nabla^{\alpha\dot\alpha}\Psi_{\alpha\beta\gamma\delta}=0$ as a
consequence of the Bianchi identities on the ASD background. Anti
self-dual spacetimes are sometimes known as `half-flat' or
`left-flat'. As discussed in~\cite{AH}, such left-flat spacetimes are all that
survive in a chiral limit of the Plebanski action, obtained by
rescaling $\Gamma\to \kappa^2\Gamma$ and then taking the limit
$\kappa^2\to0$. In this chiral theory, $\Gamma$ is
independent of the tetrad even after the field equations are imposed.

In the full theory~\eqref{gravityact}, set $\Sigma=\Sigma_0+\sigma$ and $\Gamma
=\Gamma_0+\gamma$ to consider a small fluctuation on a background
$(\Sigma_0,\Gamma_0)$. We will eventually take all fluctuations to
be proportional to the coupling $\kappa$. When the background is anti
self-dual (so $\Sigma_0$ is closed and
$\Gamma_0$ vanishes), the fluctuations are subject to the linearized field
equations
\begin{equation}
	 \rd\sigma^{\dot\alpha\dot\beta}=-2\gamma^{(\dot\alpha}_{\ \,\dot\gamma}\,
	 \Sigma_0^{\dot\beta)\dot\gamma}\qquad\hbox{and}\qquad   
	 \rd\gamma_{\dot \alpha\dot\beta}\wedge e^{\beta\dot\beta}_0=0\ .
\label{lineom}
\end{equation}
Note that the exterior derivatives $\rd$ here can be thought of as
acting covariantly on the dotted spinor indices, since
$\mathbb{S}^+\to M$ is flat in the background. After some algebra, the
second of these equations implies that \be\label{linsdweyl}
\rd\gamma_{\dot\alpha\dot\beta}=
\tilde\psi_{\dot\alpha\dot\beta\dot\gamma\dot\delta}
\Sigma_0^{\dot\gamma\dot\delta}\, , \ee where
$\tilde\psi_{\dot\alpha\dot\beta\dot\gamma\dot\delta} =
\tilde\psi_{(\dot\alpha\dot\beta\dot\gamma\dot\delta)}$. Taking the
exterior derivative of this equation and using~\eqref{graveomasd}
yields
$\nabla^{\alpha\dot\alpha}\tilde\psi_{\dot\alpha\dot\beta\dot\gamma\dot\delta}=0$,
so $\tilde\psi_{\dot\alpha\dot\beta\dot\gamma\dot\delta}$ may be
intepreted as a linearized self-dual Weyl tensor propagating on the asd
background.

Since~\eqref{lineom} are linearized, their space of solutions is a
vector space $V$. If $\cS$ is the infinite dimensional space of
solutions to the nonlinear field
equations~\eqref{graveom1}-\eqref{graveom2}, then $V$ may be thought
of as the fibre of $T\cS$ over the ASD background
$(\Sigma_0,\Gamma_0)\in\cS$. An on-shell linearized fluctuation
$(\sigma,\gamma)$ preserves the anti self-duality of the Riemann
tensor if and only if it lies in a subspace $V^-\subset V$ defined by
$\gamma^{\dot\alpha}_{\ \dot\beta}=0$, modulo gauge.  However, we cannot
invariantly define an analogous subspace $V^+$ of self-dual solutions
modulo gauge, because ({\it e.g.}) the condition that the variation
of the ASD Weyl
tensor should vanish is not true for infinitesimal diffeomorphisms and
so such a definition is not gauge invariant. (In fact, it would be
over-determined.) 
We can nevertheless define $V^+$ as the quotient $V^+= V/V^-$ so that
\begin{equation}
V^+=\{(\sigma,\gamma)\in V\}/\{(\sigma,\gamma)|\gamma^{\dot\alpha}_{\
\,\dot\beta} =\rd\mu^{\dot\alpha}_{\
\,\dot\beta}\}\,=\{\gamma\, | \rd\gamma_{\dot\alpha\dot\beta}\wedge e^{\alpha\dot\alpha}=0\}/\{\gamma^{\dot\alpha}_{\ 
\,\dot\beta} =\rd\mu^{\dot\alpha}_{\ \,\dot\beta}\}\,
. 
\label{sdlin}
\end{equation}
An element $[\sigma,\gamma]\in V^+$ determines a unique non-zero
linearized self-dual Weyl tensor by \eqref{linsdweyl}. The definitions of $V^\pm$ are summarized in
the exact sequence
\begin{equation}
0\rightarrow V^-\rightarrow V\rightarrow V^+\rightarrow 0\ ,
\label{gravseq}
\end{equation}
where the second arrow is inclusion, and the third arrow is the map
sending $(\sigma, \gamma)\rightarrow \gamma$ modulo linearized gauge
transformations. Exactness means that if a linearized solution
projects to zero in $V^+$, then it necessarily comes from one in
$V^-$.  On a flat background, $V$ decomposes as $V=V^+\oplus V^-$, but
on an ASD background such a global splitting is obstructed because
elements of $V^+$ cannot globally be required to have non-vanishing
anti self-dual parts. We will see that the MHV amplitudes precisely
measure this obstruction.

\subsection{Scattering of linearized fields}

Figure~\ref{fig:background} in the introduction realises the MHV
amplitudes as the plane wave expansion of the amplitude for the
scattering of a single, linearized graviton off an ASD background.
The linearized graviton is taken to have positive helicity in the
asymptotic past.  To fix ideas, we consider a scattering process to
take initial (characteristic) data from $\scri^-$ to data on $\scri^+$.
Here, $\scri^\pm$ are future/past null
infinity~\cite{penrose-rindler} and form the future/past boundaries
of the conformal compactification of an asymptotically flat spacetime.
They have the structure of lightcones (whose vertices are usually taken to
be at infinity), so they have topology $S^2\times \bR$.  In the
conformal compactification of Minkowski space, the lightcone of a point on $\scri^-$
refocuses on a corresponding point of $\scri^+$ and thus $\scri^\pm$ are canonically identified. The inversion
$x^a\rightarrow x^a/x^2$ sends the lightcone of the origin to
$\scri^\pm$ in the conformal compactification.

For our scattering process, the linearized graviton is prepared
to have positive helicity on $\scri^-$ and scatters off the ASD background
to emerge with negative helicity in the asymptotic future
$\scri^+$. For {\it positive frequency} fields, states of positive or
negative helicity are self-dual or anti self-dual,
respectively~\cite{birula,ashhelicity}. On a curved spacetime, one can
sometimes (perhaps with some gauge choices) define the
positive/negative frequency splitting on an arbitrary Cauchy surface,
but in general the results on different Cauchy surfaces will not
agree, as is familiar {\it e.g.}  from Hawking radiation. However, for
an asymptotically flat spacetime, $\scri^\pm$ are lightcones at
infinity' and have the same $S^2\times\bR$ topology as in Minkowski space. 
For these spacetimes, we can use
Fourier analysis in the $\bR$ factors to perform the positive/negative
frequency splitting at\footnote{Strictly, to split into
positive/negative frequency at $\scri^\pm$, we must first perform a
conformal rescaling so as to make sense of the limits of the fields
at infinity. Such conformal rescalings can be canonically restricted
to be constant along the generators~\cite{penrose-rindler} so there
is no ambiguity in the splitting.} $\scri^+$ or $\scri^-$.
Equivalently, one can split a field into parts that analytically
continue into the upper and lower half planes respectively of the
complexification $\bC$ of the $\bR$ generators.  On an asymptotically
flat spacetime that is {\it anti self-dual}, one can say more: as in
Minkowski space, the lightcone emitted from an arbitrary point of
$\scri^-$ refocusses at a point of $\scri^+$, so $\scri^\pm$ may again
be canonically identifed. (The reason for this will become transparent
in the twistor formulation of the next section; essentially,
identified points of $\scri^\pm$ correspond to the same Riemann sphere
in twistor space.)  Thus, on an ASD background, the positive/negative
frequency splittings at $\scri^-$ and $\scri^+$ agree, and it is easy
to check they reproduce the standard splitting when the spacetime is
flat. Thus we wish to find an expression for the scattering of a {\it
self-dual} linearized graviton by an arbitrary asymptotically flat,
asd spacetime $M$.

In the path integral approach, to compute the scattering amplitude, we
formally consider the integral $\int [D\Sigma D\Gamma ]\, \e^{\im
S/\hbar}$, taken over all fields that approach the prescribed
behaviour at $\scri^\pm$. In the tree-level approximation, the path
integral is given simply by evaluating $\e^{\im S/\hbar}$ on fields
that extend this boundary configuration throughout the spacetime in
accordance with the equations of motion, {\it i.e.} on
$(\Sigma_0+\sigma, \gamma)$. To leading order in the fluctuations,
this is
\begin{equation}
\e^{\im S/\hbar} \approx 1 + \frac{\im}{\kappa^2\hbar}\int_M
\Sigma_0^{\dot\alpha\dot\beta}\wedge 
\gamma^{\dot\gamma}_{\ \dot\alpha}\wedge\gamma_{\dot\gamma\dot\beta}\ .
\label{scatter}
\end{equation}
The first term on the right hand side is the diagonal part of the
S-matrix. The remaining part is the classical approximation to the
transition amplitude we seek. This term is simply $\im/\hbar$ times
the part of the Plebanski action that is lost in the chiral limit
mentioned above. Indeed, because $\Gamma $ satisfies $\rd\Gamma
_{\dot\alpha\dot\beta}\wedge e^{\beta\dot\beta}=0$ in the chiral
theory, $\Gamma$ is indistinguishable from the linearized fluctuation
in $\Gamma$ in the full theory. This field equation for $\gamma$ also
implies that the formula is gauge invariant since if we change
$\gamma\rightarrow \gamma +\rd\chi$ with $\chi$ of compact support,
the change in the integrand is clearly exact with compact support
since $\rd(\gamma _{\dot\alpha\dot\beta}\wedge
e^{\beta\dot\beta}\wedge e_\beta^{\dot\gamma})=0$ and
$\rd\Sigma_0^{\dot\alpha\dot\beta} =0$.

In the MHV diagram formulation, the
full classical theory can be built up from the complete set of MHV
vertices, together with a propagator derived from the chiral
theory\footnote{As mentioned in the Introduction, the status of the
  MHV formalism in gravity - justified using recursion relations -
  requires a more complete understanding of the possible contribution
  from the `pole at infinity'~\cite{BEF}. However, tree-level MHV
  diagrams in (super) Yang-Mills are known to be equivalent to Feynman
  diagrams~\cite{Risager,EFK}.}. Thus it is perhaps not surprising
that all of the infinite number of MHV amplitudes should somehow be
contained in this term. We will see later how to use this expression as a
generating function for all the gravitational MHV
amplitudes.

\subsubsection{An alternative derivation}
We will now rederive the expression for the scattering amplitude in
more detail.  Although this derivation is instructive, the impatient
reader may prefer to skip ahead to the next section.  Consider canonical
quantization of the Plebanski action around an anti self-dual (rather
than flat) background. The amplitude we seek might then be written as
$\langle \Phi_{\rm out}|\Phi_{\rm in}\rangle_{\rm asd}$, where
$\langle\,\cdot\,|\,\cdot\,\rangle_{\rm asd}$ is the inner product on
the Hilbert space of the theory describing fluctuations around the asd
background, and $\Phi_{\rm in}$, $\Phi_{\rm out}$ are in and out
states of the appropriate helicity.

We can construct this inner product from the symplectic form on the
phase space of the classical theory as follows (see {\it
e.g.}~\cite{woodhouse,aes}). The space of solutions $\cS$
to~\eqref{graveom1}-\eqref{graveom2} possesses a naturally defined
closed two-form $\Omega$ defined using the boundary term in the
variation of the action $S$. Letting $\delta$ denote the exterior
derivative on the space of fields, so that
$\delta\Sigma^{\dot\alpha\dot\beta}$ and $\delta\Gamma^{\dot\alpha}_{\
\dot\beta}$ are one-forms on $\cS$, $\Omega$ is given by
\begin{equation}
\Omega = \frac{1}{\kappa^2}\int_C \delta\Sigma^{\dot\alpha\dot\beta}\wedge\delta\Gamma_{\dot\alpha\dot\beta}
\label{gravsymplectic}
\end{equation}
where $C$ is a Cauchy surface in $M$.  $\Omega$ is independent of the
choice of Cauchy surface, because if $C_1$ and $C_2$ are two such surfaces bounding a region
$D\subset M$ ({\it i.e.} $\del D = C_1-C_2$) then
\begin{equation}
\int_{C_1-C_2}\hspace{-0.5cm}\delta\Sigma^{\dot\alpha\dot\beta}\wedge \delta\Gamma_{\dot\alpha\dot\beta}
=\delta\int_{\del D}\Sigma^{\dot\alpha\dot\beta}\wedge\delta\Gamma_{\dot\alpha\dot\beta} 
=\delta\int_D \rd\left(\Sigma^{\dot\alpha\dot\beta}\wedge\delta\Gamma_{\dot\alpha\dot\beta}\right)\ .
\end{equation}
Provided the field equations hold throughout $D$, this last term is
$\delta^2 S$ and so vanishes because $\delta$ is nilpotent.
Therefore, $\Omega$ is invariant under diffeomorphisms of $M$ (whether
or not these preserve $C$) and under rotations of the spin frame (it
has no free dotted spinor indices). Moreover, $\Omega$ vanishes when
evaluated on any changes in $\Sigma$ and $\Gamma $ that come from such
a diffeomorphism or spin frame rotation, so it descends to a
symplectic form on $\cS/{\rm Diff_0^+(M)}$. This symplectic form is
real for real fields in Lorentzian signature. The quantum mechanical
inner-product $\langle\,\cdot\,|\,\cdot\,\rangle$ is then defined as
\begin{equation}
\big\langle\cdot\big|\cdot\big\rangle=\frac{\im}{\hbar}\Omega(\,\cdot\, ,P_+\,\cdot\,)
\label{innerprod}
\end{equation}
where $P_+$ projects states onto their positive frequency
components\footnote{$P_+$ is a choice of `polarization' of the phase
space in which positive/negative frequency states are taken to be
holomorphic/antiholomorphic. We make this choice by defining it at
null infinity, and no ambiguity arises as to whether future or past
infinity is chosen in an asymptotically flat, ASD spacetime. One can
check that~\eqref{innerprod} is positive definite, and
linear/anti-linear in its left/right entries, with respect to the
complex structure of the polarization.}, defined at $\scri^\pm$ as
above.

We can use the symplectic form to define a duality between $V^+$ and
$V^-$.  The symplectic form vanishes on restriction to the anti
self-dual linearized solutions $V^-$ (which have $\gamma=0$, mod
gauge). So, if $h_{a,b}=(\sigma_{a,b},\gamma_{a,b})$ are two elements
of $V\simeq \left.T\cS\right|_{M_{\rm asd}}$ and $h_a \in V^-\subset
V$, then $\Omega(\,\cdot\,,h_a)$ annihilates any part of $h_b$ that is
in $V^-$ and we have
\begin{equation}\label{duality}
		\Omega(h_b,h_a) = -\frac{1}{\kappa^2}\int_C 
		\sigma_a^{\dot\alpha\dot\beta}\wedge\gamma_{b\,\dot\alpha\dot\beta}
\end{equation}
for {\it any} $(\sigma_b,\gamma_b)\in V$.  We see from this formula
that the pairing only depends on $\gamma_b$, {\it i.e.} the projection of
$(\sigma_b,\gamma_b)$ into $V^+$.  Therefore, we have an
isomorphism $V^+\buildrel{\Omega}\over{\simeq} {V^-}^*$.

We need to prepare our incoming field so that it is purely self-dual,
so we need to construct a splitting of the sequence~\eqref{gravseq}. This
is easily done on $\scri^\pm$ using the standard expression of
characteristic data for the gravitational field in terms of the
asymptotic shear $\sigma$~\cite{penrose-rindler}.  Since this expression may not familiar
to many readers, we give a somewhat formal, but
equivalent definition: motivated by \eqref{duality} we will say that a
linearized field $(\sigma_b,\gamma_b)$ is self-dual at $\scri^\pm$ if,
given a one-parameter family $C_t$ of Cauchy
hypersurfaces, with $C_t \rightarrow\scri^\pm$ as $t\rightarrow \pm\infty$, then
\be\label{sd-scri-} \lim_{t\rightarrow \pm\infty}\int_{C_t}
\sigma_b^{\dot\alpha\dot\beta}\wedge\gamma_{c\,\dot\alpha\dot\beta}=0\,
, \qquad \forall\  \gamma_c\in V^+\, .  \ee


We wish to consider the amplitude for a positive frequency, linearized
solution $h_1$ that has positive helicity at $\scri^-$ to evolve into
a positive frequency, negative helicity linearized solution at
$\scri^+$ by scattering off the ASD background. That is, $h_1$ is
purely self-dual at $\scri^-$ so it satisfies~\eqref{sd-scri-}, and
we wish to know its anti self-dual part after evolving it to
$\scri^+$. From the discussion above, we can extract this by computing
the inner product with a linearized field $h_2$ that is purely
self-dual (in ${V^-}^*$) at $\scri^+$.  Taking this inner-product at
$\scri^+$, for positive frequency states the amplitude is
\begin{equation}
\big\langle h_2\big| h_1\big\rangle = \frac\im{\hbar}\Omega(h_2,P_+h_1)
=-\frac{\im}{\kappa^2\hbar}
\int_{\scri^+}\sigma_1^{\dot\alpha\dot\beta}\wedge\gamma_{2\,\dot\alpha\dot\beta}\
\end{equation}
because $(\sigma_2,\gamma_2)$ is purely self-dual at $\scri^+$. Now,
$\del M=\scri^+-\scri^-$, so Stokes' theorem gives    
\begin{equation}
\begin{aligned}
		\big\langle h_2\big| h_1\big\rangle
		&=-\frac{\im}{\kappa^2\hbar}
		\int_M \left(\rd\sigma_1^{\dot\alpha\dot\beta}\wedge\gamma_{2\,\dot\alpha\dot\beta} 
		+ \sigma_1^{\dot\alpha\dot\beta}\wedge \rd\gamma_{2\,\dot\alpha\dot\beta}\right)
		-\frac{\im}{\kappa^2\hbar}\int_{\scri^-}\sigma_1^{\dot\alpha\dot\beta}\wedge\gamma_{2\, \dot\alpha\dot\beta}\\
		&=\frac{\im}{\kappa^2\hbar}
		\int_M\Sigma_0^{\dot\alpha\dot\beta}\wedge\gamma^{\dot\gamma}_{1\, \dot\alpha}
		\wedge\gamma_{2\,\dot\beta\dot\gamma}
		-\sigma_1^{\dot\alpha\dot\beta}\wedge\tilde\psi_{2\,\dot\alpha\dot\beta\dot\gamma\dot\delta}						\Sigma_0^{\dot\gamma\dot\delta}\\
		&=\frac{\im}{\kappa^2\hbar}
		\int_M \Sigma_0^{\dot\alpha\dot\beta}\wedge\gamma^{\dot\gamma}_{1\, \dot\alpha}\wedge
		\gamma_{2\,\dot\beta\dot\gamma}\ .
\end{aligned}
\label{background}
\end{equation}
In going to the second line, we used the linearized field
equations~\eqref{lineom} together with the fact that
$\int_{\scri^-}\sigma_2^{\dot\alpha\dot\beta}\wedge\gamma_{1\,\dot\alpha\dot\beta}=0$
because $h_1$ is purely self-dual at $\scri^-$. The third line follows
because
$\sigma^{(\dot\alpha\dot\beta}\wedge\Sigma_0^{\dot\gamma\dot\delta)}=0$
from the linearization of the constraint
$\Sigma^{(\dot\alpha\dot\beta}\wedge\Sigma^{\dot\gamma\dot\delta)}=0$
that ensures $\Sigma=\Sigma_0+\sigma$ comes from a tetrad.
Equation~\eqref{background} agrees with the form of the tree amplitude
computed before, as it should.



\section{Twistor Theory for Gravity}
\label{sec:twistors}

Although we have argued that they are related, the
expression~\eqref{scatter} (or~\eqref{background}) is still a far cry
from the usual form of the MHV amplitudes, which live on a flat
background spacetime. To connect the two pictures, we must expand out
the ASD background in~\eqref{background} in terms of plane wave
perturbations away from Minkowski space.  
This background is explicitly present
in~\eqref{background} through $\Sigma_0^{\dot\alpha\dot\beta}$ and
also implicit through the equations satisfied by the $\gamma$s. In
order to perform the expansion we will have to use the integrability
of the ASD interactions. Even so, constructing a fully nonlinear asd
background that is asymptotically a superposition of negative helicity
momentum eigenstates, and then using this background to
evaluate~\eqref{background} is a very complicated task. What enables
us to proceed is the use of twistor theory, which brings out the
integrability of the ASD sector and is therefore well-adapted to the
problem at hand.

We now briefly review the twistor theory of linearized gravity on flat
spacetime, before moving on to discuss Penrose's non-linear graviton
construction~\cite{Penrose} which gives the twistor description of an
asd spacetime (see {\it
  e.g.}~\cite{penrose-rindler,ward-wells,huggett-tod} for textbook
treatments).

\subsection{Linearized Gravity}
\label{sec:lineartwistorgrav}

We first review the basic twistor correspondence.
The twistor space $\bP\bT^\prime$ of flat spacetime is $\bC\bP^3$ with
a $\bC\bP^1$ removed. We can describe $\bC\bP^3$ using homogeneous
coordinates $Z^I=(\omega^\alpha,\pi_{\dot\alpha})$ where
$I=0,\ldots,3$, while $\alpha=0,1$ and $\dot\alpha=\dot{0},\dot{1}$
are spinor indices as before. In these coordinates, the line that is
removed is given by $\pi_{\dot\alpha}=0$, so that
$\pi_{\dot\alpha}\neq0$ on $\bP\bT^\prime$. Hence $\bP\bT^\prime$
fibres over the $\bC\bP^1$ whose homogeneous coordinates are
$\pi_{\dot\alpha}$. Points $x\in\bC^4$ of (complexified) spacetime
with coordinates $x^{\alpha\dot\alpha}$ correspond to lines
($\bC\bP^1$s) in $\bP\bT^\prime$ by the incidence relation 
\begin{equation}
\omega^\alpha=\im x^{\alpha\dot\alpha}\pi_{\dot\alpha}\ .
\label{incidence}
\end{equation}
We will denote this line by $L_x$. The removed line
$\pi_{\dot\alpha}=0$ corresponds to a point at infinity in spacetime
(the vertex of the lightcone at infinity). 

We use the standard notation $\cO(n)$ to denote the line bundle
on $\bC\bP^m$ of Chern class $n$.  Sections of $\cO(n)$ can be
identified with functions on the non-projective space of homogeneity
degree $n$, so that $Z^I\del f/\del Z^I=nf$. We will use the same
notation for line bundles over a projective line ($m=1$) and over twistor space ($m=3$).

The normal bundle to $L_x$ in $\bP\bT^\prime$ is
$N_{L_x|\bP\bT^\prime}\simeq\cO(1)\oplus\cO(1)$.  In particular, for
$x=0$, $\omega^\alpha$ are coordinates along the fibres of the normal
bundle to $L_0$. Thus, in this flat case, $\bP\bT^\prime$ is the total
space of the normal bundle to a line. The incidence
relation~(\ref{incidence}) identifies a point $x$ with a holomorphic
section $\bC\bP^1\to\bP\bT^\prime$ and the space of such sections
$H^0(L_x,N_{L_x|\bP\bT^\prime})\simeq\bC^4$ is (complexified) flat
spacetime.

The correspondence with flat spacetime can also be expressed
in terms of the double fibration 
\begin{equation}
\begin{aligned}
\begin{picture}(50,40)
\put(0.0,0.0){\makebox(0,0)[c]{$\bP\bT^\prime$}}
\put(57.0,0.0){\makebox(0,0)[c]{$\bM$}}
\put(34.0,33.0){\makebox(0,0)[c]{$P(\mathbb{S}^+)$}}
\put(7.0,18.0){\makebox(0,0)[c]{$p$}}
\put(55.0,18.0){\makebox(0,0)[c]{$q$}}
\put(25.0,25.0){\vector(-1,-1){18}}
\put(37.0,25.0){\vector(1,-1){18}}
\end{picture}
\end{aligned}
\label{doublefibration}
\end{equation}
where $P(\mathbb{S}^+)$ is the projectivization of the bundle of
dotted spinors, coordinatized by $(x^{\alpha\dot\alpha},
\pi_{\dot\beta})$ up to scaling of the $\pi$s, and $\bM\simeq\bC^4$ is complexified Minkowski space. The bundle
$P(\mathbb{S}^+)\to \bM$ is necessarily trivial, and the fibres
$q^{-1}(x)$ are $\bC\bP^1$s coordinatized by
$\pi_{\dot\alpha}$. Conversely, the fibres
$p^{-1}(\omega^\alpha,\pi^{\dot\alpha})$ are the set of points
$(x^{\alpha\dot\alpha},\pi_{\dot\alpha})$ such that $\omega^\alpha=\im
x^{\alpha\dot\alpha}\pi_{\dot\alpha}$; given one such point
$(x_0,\pi)$, this is the totally null, complex two-plane
$x_0^{\alpha\dot\alpha}+\lambda^\alpha\pi^{\dot\alpha}$.

\smallskip

The Penrose transform represents linearized gravitons of helicities
$-2$ and $+2$ on spacetime as elements of the twistor space cohomology groups
$H^1(\bP\bT^\prime,\cO(2))$ and $H^1(\bP\bT^\prime,\cO(-6))$,
respectively. In a Dolbeault framework, these are
described locally by $(0,1)$-forms $h(Z)$ and $\tilde h(Z)$,
homogeneous of degrees $2$ and $-6$. $h$ and $\tilde h$ thus obey $\delbar
h=0=\delbar\tilde h$ and are defined up to the gauge freedom $h\sim h+\delbar\chi$, 
$\tilde h\sim\tilde h+\delbar\lambda$. We will suppress $(0,p)$-form indices in what follows (and some readers may prefer to think in terms of a {\v C}ech picture of cohomology). The Penrose transforms of $h$ and $\tilde h$ are 
\begin{equation}
\begin{aligned}
		\psi_{\alpha\beta\gamma\delta}(x)&=\int_{L_x}[\pi\,\rd\pi]\wedge
		p^*\left(\frac{\del^4 h}{\del\omega^\alpha\del\omega^\beta\del\omega^\gamma\del\omega^\delta}
		\right)\\
		\tilde\psi_{\dot\alpha\dot\beta\dot\gamma\dot\delta}(x)
		&=\int_{L_x} [\pi\,\rd\pi] \wedge \pi_{\dot\alpha}\pi_{\dot\beta}\pi_{\dot\gamma}\pi_{\dot\delta}\,
		p^*(\tilde h)
\end{aligned}
\label{htrans}
\end{equation}
where the pullback $p^*$ simply imposes the incidence relation~(\ref{incidence}). Differentiating under the integral sign shows that $\psi$ and $\tilde\psi$ obey the usual spin-2 ({\it i.e.} linearized Einstein) equations
$\del^{\alpha\dot\alpha}\psi_{\alpha\beta\gamma\delta}=0,\ \
\del^{\alpha\dot\alpha}\tilde
\psi_{\dot\alpha\dot\beta\dot\gamma\dot\delta}=0$ provided only that $h$
and $\tilde h$ are $\delbar$-closed.

The cohomology class $h$ plays an active role through its associated
Hamiltonian vector field
\begin{equation}
V := I(\rd h,\,\cdot\,) = I^{JK}\frac{\del h}{\del Z^J}\frac{\del\ }{\del Z^K}\ .
\label{Vdef}
\end{equation}
Here $I$ is a holomorphic Poisson bivector of homogeneity $-2$. It is determined by the line that was removed from $\bC\bP^3$ to reach $\bP\bT^\prime$ and has components
\begin{equation}
		I^{JK}=
		\begin{pmatrix}
		\varepsilon^{\alpha\beta}&0\\
		0&0\\
		\end{pmatrix}\qquad \hbox{so that}\qquad
		I = \varepsilon^{\alpha\beta}\frac{\del\ }{\del\omega^\alpha}\wedge\frac{\del\ }{\del\omega^\beta}\ .
\label{infinity}
\end{equation}
It follows that $V$ in~(\ref{Vdef}) represents an element of $H^1(\bP\bT^\prime,T_{\bP\bT^\prime})$ and so describes a linearized complex structure deformation. We will study these deformations further in the next subsection.

A positive helicity graviton may also be represented by an element
\be B\in H^1(\bP\bT^\prime, \Omega^{1,0}\otimes\cO(-4))\ee if, as well as
having the standard gauge freedom $B\to B+\delbar \chi$ of a
cohomology class, $B$ is also subject to the additional gauge freedom
\begin{equation}
		B\to B+\del m + n[\pi\,\rd\pi]\ .
\label{Bgauge}
\end{equation}
Here, $m$ and $n$ are $(0,1)$-forms of homogeneity $-4$ and $-6$ respectively, while $\chi$ is a $(1,0)$-form of
weight $-4$. The freedom to add on arbitrary multiples of $[\pi\,\rd\pi]$ means that only the part $B_\alpha \rd\omega^\alpha$ of $B$ along the fibres of $\bP\bT^\prime\to\bC\bP^1$ contains physical information; the remaining freedom 
$B\to B+\del m$ means that this physical information is captured by
\begin{equation}
		I(dB) = I^{IJ}\del_I B_J = \varepsilon^{\alpha\beta}\frac{\del B_\beta}{\del\omega^\alpha}\ .
\label{Bdef}
\end{equation}
$I(dB)$ is again in $H^1(\bP\bT^\prime,\cO(-6))$ and so can be identified with $\tilde h$. The Penrose transform of $B$ is
\begin{equation}
		\gamma^{\dot\alpha}_{\ \dot\beta} =2\int_{L_x}[\pi\,\rd\pi]\wedge\pi^{\dot\alpha}\pi_{\dot\beta}\,  p^*(B)
\label{Btrans}
\end{equation}
which, as our notation suggests, may be interpreted as a linearized self-dual spin connection. (The factor of 2 is for later convenience.) To see this, note first that~(\ref{Btrans}) respects the gauge freedom~(\ref{Bgauge}) because
any piece of $p^*B$ proportional to $[\pi\,\rd\pi]$ wedges to zero
in~(\ref{Btrans}), while adding on a total derivative $B\to B+dm$
corresponds to the linearized gauge freedom
$\gamma^{\dot\alpha}_{\ \dot\beta}\to\gamma^{\dot\alpha}_{\ \dot\beta} + \rd\mu^{\dot\alpha}_{\ \dot\beta}$ of a spacetime connection. ($\mu^{\dot\alpha}_{\ \dot\beta}$ is the Penrose transform of $m$ and satisfies the
asd Maxwell equation $\del_{\alpha\dot\alpha}\mu^{\dot\alpha}_{\ \dot\beta}=0$.) The linearized spin connection generates a linearized curvature fluctuation as it ought, since 
\begin{equation}
\begin{aligned}
		\rd\gamma_{\dot\alpha\dot\beta}
		&=2dx^{\delta\dot\delta}\frac{\del\ }{\del x^{\delta\dot\delta}}
		\left(\int_{L_x}[\pi\,\rd\pi]\wedge\pi_{\dot\alpha}\pi_{\dot\beta}\, p^*(B)\right)\\
		&=2dx^{\delta\dot\delta}\wedge dx^{\gamma\dot\gamma}
		\int_{L_x}[\pi\,\rd\pi]\wedge\pi_{\dot\alpha}\pi_{\dot\beta}\pi_{\dot\gamma}\pi_{\dot\delta}\,
		p^*\left(\frac{\del B_\gamma}{\del\omega^\delta}\right)\\
		&=dx^{\delta\dot\delta}\wedge dx_\delta^{\ \dot\gamma}
		\int_{L_x}[\pi\,\rd\pi]\wedge\pi_{\dot\alpha}\pi_{\dot\beta}\pi_{\dot\gamma}\pi_{\dot\delta}\,
		p^*\!\left(\tilde h\right)\\
		&=\tilde\psi_{\dot\alpha\dot\beta\dot\gamma\dot\delta}\,dx^{\delta\dot\delta}\wedge dx_\delta^{\ \dot\gamma}
\end{aligned}
\end{equation}
where in the second line we used the fact that 
$p^*B=\im B_\gamma(\im x\!\cdot\!\pi,\pi)dx^{\gamma\dot\gamma}\pi_{\dot\gamma}$ (mod $[\pi\,\rd\pi]$), which depends on $x$ only through $\omega^\gamma=\im x^{\gamma\dot\gamma}\pi_{\dot\gamma}$.

Plane wave
gravitons (linearized spin-2 fields) of momentum
$p_{\alpha\dot\alpha}=\tilde k_\alpha k_{\dot\alpha}$ may be described
by twistor functions
\begin{equation}
h(Z) = \kappa\,\bar\delta_{(2)}([\pi \, k])\exp\left(\langle\omega\, \tilde k\rangle\right)\qquad\qquad
\tilde{h}(Z) = \kappa\,\bar\delta_{(-6)}([\pi \,k])\exp\left(\langle\omega \, \tilde k\rangle\right)\ .
\label{planewave}
\end{equation}
where, for later use, we have taken all fluctuations to be
proportional to the coupling $\kappa=\sqrt{16\pi{\rm G_N}}$ and we
follow~\cite{CSW,BMS2} in defining
\begin{equation}
\bar\delta_{(r)}\left([\pi\,k]\right):=\left(\frac{[\pi\,\alpha]}{[k\,\alpha]}\right)^{r+1}
\delbar\frac{1}{[\pi\,k]}\ .
\label{deltabardef}
\end{equation}
In this definition, $|\alpha]$ is a fixed dotted spinor introduced so
that the $\delta$-function (0,1)-forms
$\bar\delta_{(r)}\left([\pi\,k]\right)$ have homogeneity $r$ in
$|\pi]$. On the support of the $\bar\delta$-function,
$\pi_{\dot\alpha}\propto k_{\dot\alpha}$ so the momentum
eigenstates~\eqref{planewave} are in fact independent of the choice of
$|\alpha]$. Note that, because of the weight of the
$\bar\delta$-function, $h$ has weight $-4$ in the momentum spinor
$|k]$ (counting $|\tilde k\rangle$ as weight $-1$), while $\tilde h$
has weight $+4$. This is as expected for states of helicity $-2$ and
$+2$, respectively.

Likewise, the one-forms $B$ may be taken to be 
\be\label{Bplanewave} 
		B(Z) =\kappa\frac{\langle\tilde\beta\,\rd\omega\rangle}{\langle\tilde\beta\,\tilde k\rangle}
			\bar\delta_{(-5)}\left([\pi\,k]\right)\,\exp\left(\langle\omega\,\tilde k\rangle\right)\  ,
\ee
where the constant
undotted spinor $\langle\tilde\beta|$ arises from the gauge
freedom~(\ref{Bgauge}) in the definition of $B$. The choice of
$\langle\tilde\beta|$ is arbitrary provided
$\langle\tilde\beta\,\tilde k\rangle\neq0$ reflecting the gauge
freedom \eqref{Bgauge}. 
It is easy to check that
$I(dB)=\tilde h$, with $\tilde h$ as above in \eqref{planewave}.

\smallskip

We remark in passing that $V$ represents an element of
$H^1(\bP\bT^\prime,T_{\bP\bT^\prime})$ together with the extra
requirement~(\ref{Vdef}) that it be Hamiltonian with respect to $I$,
while (incorporating the redundancy $B\to B + \delbar\chi+\del m$) $B$
represents an element of $H^1(\bP\bT^\prime,\Omega^2_{\rm
cl}\otimes\cO(-4))$ where $\Omega^2_{\rm cl}$ is the sheaf of closed
(2,0)-forms, together with the extra requirement that $n[\pi\,\rd\pi]$
be taken equivalent to zero. Without the Hamiltonian and
$n[\pi\,\rd\pi]\sim0$ conditions, these cohomology groups represent
states in conformal gravity. The extra conditions eliminate half the
conformal gravity spectrum, reducing it to Einstein gravity as
above. The cohomology groups $H^1(\bP\bT^\prime,T_{\bP\bT^\prime})$
and $H^1(\bP\bT^\prime,\Omega^2_{\rm cl}\otimes\cO(-4))$ (together
with their $\cN=4$ completions) define vertex operators in the Witten,
Berkovits or heterotic twistor-string theories~\cite{BW,het}.  String
theories that impose the extra conditions were constructed
in~\cite{AHM}, but these theories only seem to describe the asd
interactions of Einstein (super)gravity~\cite{Nair2}.

\subsection{The Non-Linear Graviton}
\label{sec:nonlineargrav}

Penrose's non-linear graviton construction~\cite{Penrose} associates a
deformed twistor space $\cP\cT$ to a spacetime ${M}$ with anti
self-dual (ASD) Weyl tensor. In this correspondence, the structure of
$M$ is encoded into the deformed complex structure of the twistor
space.  For ASD spacetimes that also obey the vacuum Einstein equations,
the twistor space fibres over $\bC\bP^1$ and admits an analogue of the
Poisson structure $I$ along the fibres. We can still describe such a
$\cP\cT$ using homogeneous coordinates
$(\omega^\alpha,\pi_{\dot\alpha})$, where $\pi_{\dot\alpha}$ are
holomorphic coordinates that are homogenous coordinates for the
$\bC\bP^1$ base. As in $\bP\bT^\prime$, $\omega^\alpha$ parametrize
the fibres of $\cP\cT\to\bC\bP^1$, but in general they will no longer
be holomorphic coordinates throughout the deformed twistor space. As
in flat space, ${M}$ is reconstructed as the space of degree-1
holomorphically embedded $\bC\bP^1$s inside $\cP\cT$. For some fixed
$x\in {M}$, we will again denote the corresponding $\bC\bP^1$ by
$L_x$.  Although it will no longer have all the properties of a
`straight line', the normal bundle $N_{L_x|\cP\cT}$ will still be
$\cO(1)\oplus\cO(1)$ (as it was in the flat case) so that
$H^0(L_x,N_{L_x|\cP\cT})\simeq\bC^4$, which is identified as the {\it
tangent} space $\left.T{M}\right|_x$.  Just as spacetime is no
longer an affine vector space, $\cP\cT$ is no longer isomorphic to the
total space of $N_{L_x|\cP\cT}$. The correspondence may again be
interpreted in terms of a double fibration
\begin{equation}
\begin{aligned}
		\begin{picture}(50,40)
		\put(0.0,0.0){\makebox(0,0)[c]{$\cP\cT$}}
		\put(57.0,0.0){\makebox(0,0)[c]{${M}$}}
		\put(34.0,38.0){\makebox(0,0)[c]{$P(\mathbb{S}^+)$}}
		\put(7.0,20.0){\makebox(0,0)[c]{$p$}}
		\put(55.0,20.0){\makebox(0,0)[c]{$q$}}
		\put(25.0,26.0){\vector(-1,-1){18}}
		\put(37.0,26.0){\vector(1,-1){18}}
\end{picture}
\end{aligned}
\label{doublefibration2}
\end{equation}
as in~(\ref{doublefibration}). For a half-flat spacetime ${M}$ that is
sufficiently close to flat spacetime $\bM$, the spin bundle is the
product $\bC\bP^1\times {M}$.

The complex structure on $\cP\cT$ may be described in terms of a
finite deformation of the flat background $\delbar$-operator:
\begin{equation}
		\delbar\to \delbar+V=\delbar + I(\rd h,\,\cdot\,)
\label{delbardef}
\end{equation}
with $I(dh,\,\cdot\,)$ as in equation~(\ref{Vdef}).  Only allowing Hamiltonian deformations of the $\delbar$-operator ensures that $\cP\cT$ also fibres over $\bC\bP^1$ and has a holomorphic Poisson structure $I^\prime$ on the fibres. This will be essential in the construction of the spacetime metric below. The deformed
$\delbar$-operator defines an integrable almost complex structure if and only if
the Nijenhuis tensor 
\begin{equation}
\label{Nijenhuis} 
		N=(\delbar +V)^2\in\Omega^{0,2}(\bP\bT^\prime,T_{\bP\bT^\prime})
\end{equation}
vanishes. For Hamiltonian deformations~(\ref{delbardef}), one finds~\cite{MW} $N=0$ if 
\begin{equation}
\label{Nijenhuis-MW}
		\delbar h + \frac{1}{2}\{h,h\}=0\ .
\end{equation}
There is a `Poisson diffeomorphism' freedom generated by Hamiltonians $\chi$
which are smooth functions of weight two, because changing
\begin{equation}
		h\to h + \delbar\chi + \{h,\chi\}
\label{Poissondiff}
\end{equation}
does not alter the complex structure. 
The diffeomorphism freedom can be fixed by requiring $h$ to be holomorphic in $\omega^\alpha$ and proportional to 
$\langle\bar\pi\,\rd\bar\pi\rangle$, so that its (0,1)-form is purely along the base of the fibration $\cP\cT\to\bC\bP^1$. Any
such $h$ automatically leads to a vanishing Nijenhuis tensor. This gauge condition is natural in a scattering theory 
context, being essentially the same condition as is utilised in Newman's formulation of the nonlinear 
graviton~\cite{Newman,HNPT,ET}. In Newman's formulation (which will
not be emphasized here), the holomorphic lines $L_x$ are obtained from
lightcone cuts of (complexified) null infinity $\bC\scri$ and can thus
be reconstructed simply from the asymptotic data of the spacetime $M$,
while $h$ is interpreted as an integral of the asymptotic shear (the
asymptotic characteristic data of $M$). Requiring that $h$ be
holomorphic in $\omega^\alpha$ and proportional to
$\langle\bar\pi\,\rd\bar\pi\rangle$ does not completely fix the gauge
freedom~\eqref{Poissondiff}. In Newman's picture, the remaining
freedom is fixed by additionally requiring that $h$ depends on
$\omega^\alpha$ only through $\langle\omega\,\bar\pi\rangle$. We will
implicitly use `Newman gauge' in what follows: in particular, the
twistor representatives of momentum eigenstates introduced in
equation~\eqref{planewave} are adapted to Newman gauge.

\smallskip

As mentioned above, each point $x\in{M}$ corresponds to a 
holomorphically embedded $\bC\bP^1$ denoted by $L_x$. The flat space incidence relation $\omega^\alpha=\im x^{\alpha\dot\alpha}\pi_{\dot\alpha}$ must be generalized, because $\omega^\alpha$ is no longer a globally holomorphic coordinate on $\cP\cT$. We thus represent $L_x\subset\cP\cT$ by the deformed incidence relation
\begin{equation}
\omega^\alpha=F^\alpha(x,\pi)
\label{incidef}
\end{equation}
where $F^\alpha$ has homogeneity one in $\pi_{\dot\alpha}$.  The
condition that $L_x$ be holomorphic with respect to the deformed
complex structure~(\ref{delbardef}) is
\begin{equation}
		0=\left.(\delbar+V)(\omega^\alpha-F^\alpha(x,\pi))\right|_{L_x}
		=\left.V^\alpha\right|_{L_x}-\delbar F^\alpha(x,\pi)\ ,
\end{equation}
so that we obtain the condition
\be\label{Fsolv}
\delbar F^\alpha(x,\pi) = V^\alpha(F^\alpha(x,\pi), \pi)\, .
\ee
The restriction of $V^\alpha$ to $L_x$ means that we set
$\omega^\alpha=F^\alpha(x,\pi)$ in $V$, so that~(\ref{Fsolv}) is a
nonlinear differential equation for $F^\alpha$.  This generally makes
it very difficult to find explicit expressions for the holomorphic
curves. As in $\bP\bT^\prime$, for fixed $x$ the curve
$L_x\subset\mathcal{PT}$ defined by~(\ref{Fsolv}) is a section of the
fibration $\cP\cT\to\bC\bP^1$, holomorphic with respect to the
deformed complex structure, and has normal bundle
$N_{L_x|\cP\cT}\simeq\cO(1)\oplus\cO(1)$. The 
deformation theory of Kodaira \& Spencer implies that the family of
lines in $\bP\bT^\prime$ survive small deformations of the complex
structure and form a four parameter family. Thus 
there will be a four parameter space of solutions to the
nonlinear equation~(\ref{Fsolv}) and it is this parameter space that
we identify with ${M}$.  

\begin{figure}[t]
\begin{center}
\includegraphics[height=45mm]{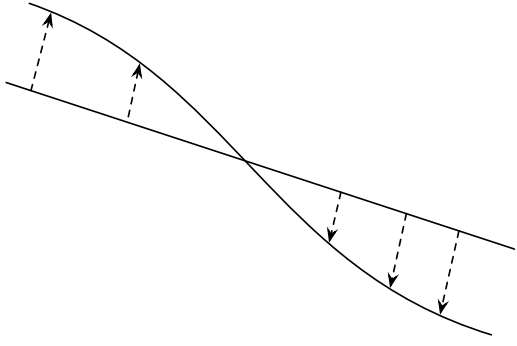}
\caption{{\it Deformations of the complex structure induce deformations of
the holomorphic curves. Identifying the four parameters $x$ on which
$F^\alpha(x,\pi)$ depends with spacetime coordinates, the normal
vector $\left(F^\alpha-\im x^{\alpha\dot\alpha}\pi_{\dot\alpha}\right)\del/\del\omega^\alpha$
on $L_x$ connects the original twistor line $\omega^\alpha=\im
x^{\alpha\dot\alpha}\pi_{\dot\alpha}$ to the deformed curve.}}
\label{fig:deform}
\end{center}
\end{figure}

\subsubsection{Constructing the spacetime metric}

The space of degree one curves is naturally endowed with a conformal structure
by requiring two points $x,y\in {M}$ to be connected by a null
geodesic if their corresponding `lines' $L_x,L_y\subset\cP\cT$
intersect.  Let us now show explicitly how to use the twistor data to construct a spacetime metric~\cite{Penrose}.

Consider the (weighted)
1-forms $[\pi\,\rd\pi]$ and $\rd\omega^\alpha-V^\alpha$. These forms are annihilated by contraction with the
antiholomorphic vector fields of the deformed complex structure, and
so define a basis of holomorphic forms on $\cP\cT$. Note that the holomorphic form $[\pi\,\rd\pi]$ is unaltered compared to $\bP\bT^\prime$; this is a consequence of restricting to Hamiltonian complex structure deformations in~\eqref{delbardef}. The holomorphic
3-form of weight $+4$ is therefore
$\Omega_{\cP\cT}=[\pi\,\rd\pi]\wedge(\rd\omega^\alpha-V^\alpha)\wedge
(\rd\omega_\alpha-V_\alpha)$. Pulling  
back $\Omega_{\cP\cT}$ to $P(\mathbb{S}^+)$ ({\it i.e.} imposing the
incidence relation $\omega^\alpha=F^\alpha(x,\pi)$) gives
\begin{equation}
\begin{aligned}
		p^*\Omega_{\cP\cT}
		&=[\pi\,\rd\pi]\wedge p^*(\rd\omega^\alpha-V^\alpha)\wedge
		p^*(\rd\omega_\alpha-V_\alpha)\\
		&=[\pi\,\rd\pi]\wedge \rd_x  F^\alpha\wedge \rd_x  F_\alpha
\end{aligned}
\label{pullback}
\end{equation}
where $\rd_x$ denotes the exterior derivative on  $P(\mathbb{S}^+)$
holding $\pi_{\dot\alpha}$ constant, {\it i.e.} $\rd_x=\rd x^a\del/ \del x^a$. (Possible terms in $\rd\bar\pi$ vanish by virtue of
the holomorphy of these sections, while terms in $\rd\pi$ vanish by virtue of
the fact that the expressions are wedged against $ [\pi\,\rd\pi]$.)
The requirement~(\ref{Fsolv}) that $L_x\subset\mathcal{PT}$ is a
holomorphic line ensures 
$\delbar F^\alpha(x,\pi)=V^\alpha(F(x, \pi), \pi)$ so
\begin{equation}
		\delbar\left( \rd_x F^\alpha\wedge \rd_x F_\alpha\right)
		= 2\,\rd_x(\delbar F^\alpha)\wedge \rd_x F_\alpha
		= 2\left.\del_{\beta} V^\alpha\right|_{L_x}\rd_x F^\beta\wedge \rd_x F_\alpha
\end{equation}
where in the second term we used the fact that $V^\alpha$ depends on $x$ only through $F^\beta(x,\pi)$. The wedge product implies this expression is antisymmetric in $\alpha,\beta$ and so in fact it vanishes because $V$ is
Hamiltonian. Therefore $\rd_x F^\alpha\wedge \rd_x F_\alpha$ is a two-form
of homogeneity $+2$ in $\pi_{\dot\alpha}$ that is holomorphic along each
$\bC\bP^1$. Consequently, by Liouville's theorem,
\begin{equation}
p^*\Omega_{\mathcal{PT}} =
    -[\pi\,\rd\pi]\wedge q^*\Sigma^{\dot\alpha\dot\beta}(x)\pi_{\dot\alpha}\pi_{\dot\beta} 
\label{sigmadef}
\end{equation}
where $\Sigma^{\dot\alpha\dot\beta}\in\Omega^2({M},{\rm Sym}^2\,\mathbb{S}^+)$ are three spacetime two-forms, pulled back to $P(\mathbb{S}^+)$ by $q^*$. (The minus sign is for convenience.) We drop the pullback symbol $q^*$ in what follows.

$\Sigma^{\dot\alpha\dot\beta}$ is automatically closed on spacetime, because $\Sigma^{\dot\alpha\dot\beta}\pi_{\dot\alpha}\pi_{\dot\beta}=\rd_x F^\alpha\wedge\rd_x F_\alpha$.  The discussion around equation~\eqref{graveomasd} then shows that the spacetime ${M}$ is {\it necessarily anti self-dual}. Moreover, $\Sigma^{\dot\alpha\dot\beta}$ is {\it simple} by construction, so 
\begin{equation}
		\pi_{\dot\alpha}\pi_{\dot\beta}\Sigma^{\dot\alpha\dot\beta}
			=\pi_{\dot\alpha}e^{\alpha\dot\alpha}\wedge\pi_{\dot\beta} e_\alpha^{\ \dot\beta}\ .
\label{sigmasimple}
\end{equation}
for some tetrad $e^{\alpha\dot\alpha}$. This decomposition does not uniquely fix the tetrad: we can freely replace $e^{\alpha\dot\alpha}$ by  $\Lambda^{\alpha}_{\ \beta}e^{\beta\dot\alpha}$ for $\Lambda^{\alpha}_{\ \beta}(x,\pi)$ an arbitrary element of 
$SL(2,\bC)$, as any such $\Lambda^\alpha_{\ \beta}$ drops out of equation~\eqref{sigmasimple}. Comparing definitions shows that
\begin{equation}
\begin{aligned}
		p^*(\rd\omega^\alpha-V^\alpha)&=\rd_x F^\alpha \qquad\qquad&\hbox{mod}\ [\pi\,\rd\pi]&\\
		&=\im\Lambda^\alpha_{\ \beta}\,e^{\beta\dot\beta}\pi_{\dot\beta}&\hbox{mod}\ [\pi\,\rd\pi]&\ .
\end{aligned}
\label{edef}
\end{equation}
Equations~(\ref{sigmasimple}) \&~(\ref{edef}) generalize the flat
spacetime formul\ae\footnote{Strictly, equations~\eqref{flatvierbein}
also includes a $\Lambda^\alpha_{\ \beta}$ in the definition of
$p^*\rd\omega^\alpha$. Such a $\Lambda$ relates the twistor
coordinate index on $\omega^\alpha$ to the undotted
spacetime spinor index on $dx^{\alpha\dot\alpha}$. On a flat
background these indices can be identified directly.}
\begin{equation}
\begin{aligned}
	p^*(\rd\omega^\alpha\wedge \rd\omega_\alpha)
	&=-\rd x^{\alpha\dot\alpha}\wedge \rd x_\alpha^{\
      \dot\beta}\,\pi_{\dot\alpha}\pi_{\dot\beta}\qquad  
	&\hbox{mod }[\pi\,\rd\pi]\\
	p^*\rd\omega^\alpha&=\im\,\rd x^{\alpha\dot\alpha}\pi_{\dot\alpha}
    &\hbox{mod }[\pi\,\rd\pi] 
\end{aligned}
\label{flatvierbein}
\end{equation}
arising from the incidence relation $\omega^\alpha={\rm i}x^{\alpha\dot\alpha}\pi_{\dot\alpha}$ in $\mathbb{PT}^\prime$.

In~\eqref{edef}, a choice of $\Lambda^\alpha_{\ \beta}$ fixes a choice of spin frame (for the undotted spinors) and hence a choice of tetrad $e^{\alpha\dot\alpha}$. However, although $\Lambda^\alpha_{\ \beta}(x,\pi)$ has weight zero in 
$\pi_{\dot\alpha}$, generically it is not $\pi$-independent. Because of this, it is not simply a local Lorentz transform on spacetime, but is best thought of as a holomorphic frame\footnote{$\Lambda^\alpha_{\ \beta}(x,\pi)$ is thus somewhat analogous to the choice of holomorphic frame $H(x,\pi)$ that arises in a similar context for Yang-Mills, see equations~\eqref{Sparling}-\eqref{A-def}.} trivializing $N_{L_x|\cP\cT}\otimes \cO(-1)$ over $L_x$ (see also~\cite{CDM,MasonFrauen}). Note that since the normal bundle $N_{L_x|\cP\cT}\simeq\cO(1)\oplus\cO(1)$, the bundle $N_{L_x|\cP\cT}\otimes \cO(-1)$ is indeed trivial on $L_x$. Its space of global holomorphic sections 
$H^0(L_x,\cO\oplus\cO)\simeq\bC^2$ is precisely the fibre $\left.\mathbb{S}^-\right|_x$ of the bundle of {\it anti} self-dual spinors on ${M}$.


\section{Gravitational MHV amplitudes from twistor space}
\label{sec:twistamps}

We now provide a twistorial description of $\langle h_2|h_1\rangle$ by
translating the right hand side of~(\ref{background}) using the
Penrose integral transform. Finally, we will use the twistor
description to expand around Minkowski space in plane waves, thus
recovering the standard form of the MHV amplitudes.  Underlying much
of what follows is a presentation for the twistor data, going back to
Newman~\cite{Newman}, that relates directly to the asymptotic data at
$\scri$ for the fields involved.  By using sums of momentum
eigenstates for the data at $\scri$ we guarantee that the fields and
backgrounds that we work with are {\it asymptotically} superpositions
of plane waves.  Technically, ASD spacetimes constructed in this way
are not asymptotically flat along the directions of the plane waves.
It is nevertheless possible to incorporate them into an asymptotically
flat formalism at the expense of having to consider $\delta$-function
singularities in the asymptotic data (the asymptotic shear) as already
apparent in \eqref{planewave} for the twistor representatives.

\smallskip

In section~\ref{sec:asd}, the classical amplitude for a positive
helicity graviton to cross an asymptotically flat ASD spacetime and
emerge with negative helicity was shown to be
\begin{equation}
\big\langle h_n\big| h_1\big\rangle 
= \frac{\im}{\kappa^2\hbar}\int_{{M}}\Sigma_0^{\dot\alpha\dot\beta}\wedge
\gamma^{\ \dot\gamma}_{n\ \dot\alpha}\wedge\gamma_{1\,\dot\beta\dot\gamma}\ ,
\label{reminder}
\end{equation}
where $\Sigma_0^{\dot\alpha\dot\beta}$ is formed from the tetrad of
the half-flat background and $\gamma_{1,n}$ are two linearized
self-dual connections that are on-shell with respect to the linearized
field equations~\eqref{lineom}. (The labelling $1,n$ is for later
convenience.) We seek a twistorial interpretation of this term.

Firstly, the Penrose transform~(\ref{Btrans}) of the linearized self-dual spin connection 1-form
\begin{equation}
		\gamma^{\dot\alpha}_{\ \dot\beta}
		=2\int_{L_x}[\pi\,\rd\pi]\wedge\pi^{\dot\alpha}\pi_{\dot\beta}\,p^*\left(B\right)
\label{Btranscurved}
\end{equation}
also makes sense on an ASD background. To see
this, first recall from section~\ref{sec:asd} that the background self-dual spin connection is flat on an ASD spacetime. It is therefore at most pure gauge and can be taken to vanish. The space of dotted spinors is then globally trivialized both on spacetime and on twistor space, so there is no difficulty in adding
$\pi^{\dot\alpha}\pi_{\dot\beta}$ at different points of $L_x$
in~(\ref{Btranscurved}). As in equation~(\ref{Btrans}) for flat space,
the Penrose transform~(\ref{Btranscurved}) is the pullback of the
$(2,1)$-form $[\pi\,\rd\pi]\wedge\pi^{\dot\alpha}\pi_{\dot\beta}B$ to
$P(\mathbb{S}^+)$, pushed down to the ASD spacetime ({\it i.e.} integrated
over the $\bC\bP^1$ fibres of $P(\mathbb{S}^+)\buildrel{q}\over{\to} {M}$). To see
that this pushdown is well-defined, note that for any vector field $X$
on ${M}$, there is a unique vector field $\tilde X\in
TP(\mathbb{S}^+)$ that obeys $\tilde X\,\lrcorner\,[\pi\,\rd\pi]=0$ and whose projection $q_*(\tilde X)$ to $T{M}$ is 
again $X$. So for any such $X$, the integral $2\int_{L_x}\tilde
X\,\lrcorner\,\left([\pi\,\rd\pi]\wedge\pi^{\dot\alpha}\pi_{\dot\beta}\,p^*B\right)$
is well-defined and equal to $X\,\lrcorner\,\gamma^{\dot\alpha}_{\ \dot\beta}$. Hence the integral~(\ref{Btranscurved}) is also unambiguous. In particular, if $\{\nabla_{\gamma\dot\gamma}\}$ is a
basis of $T{M}$ dual to the tetrad, the components of the spin
connection in this basis are given by contracting~(\ref{Btranscurved})
with $\nabla_{\gamma\dot\gamma}$:
\begin{equation}
		(\gamma_{\gamma\dot\gamma})^{\dot\alpha}_{\ \dot\beta}
		:=\nabla_{\gamma\dot\gamma}\,\lrcorner\,\gamma^{\dot\alpha}_{\ \dot\beta}
		=2\int_{L_x}[\pi\,\rd\pi]\wedge\pi^{\dot\alpha}\pi_{\dot\beta}
		\pi_{\dot\gamma} \,B_\alpha(F,\pi)\Lambda^\alpha_{\ \gamma}(x,\pi)\ ,
\label{Bcompstrans}
\end{equation}
where we have used~(\ref{edef}) to evaluate
$\tilde\nabla_{\gamma\dot\gamma}\,\lrcorner\,p^*B$.  The holomorphic frame
$\Lambda^{\alpha}_{\ \beta}$ trivializes the anti self-dual spin bundle over $L_x$, thus allowing us to makes sense of the integral of the indexed quantity\footnote{A similar r\^ole is played by the holomorphic frame $H$ in the Penrose transform of a background coupled self-dual Yang-Mills field, see~\eqref{Gtrans}.} $B_\alpha$.

To construct the Penrose transform of the expression for $\langle h_n|h_1\rangle$, we extract the components of each $\gamma$ to obtain
\begin{equation}
		\frac{\im}{\kappa^2\hbar}\int_{{M}}\Sigma_0^{\dot\alpha\dot\beta}\wedge
		\gamma^{\ \dot\gamma}_{n\ \dot\alpha}\wedge\gamma_{1\,\dot\beta\dot\gamma}
		=\frac{\im}{2\kappa^2\hbar}\int_{M} \rd\mu\ \gamma_n^{\gamma\dot\gamma\dot\alpha\dot\beta}
		\gamma_{1\,\gamma\dot\gamma\dot\alpha\dot\beta}\ ,
\end{equation}
where $\rd\mu:=\Sigma_0^{\dot\alpha\dot\beta}\wedge\Sigma_{0\,\dot\alpha\dot\beta}$ is the volume form on ${M}$. Using the Penrose transform~(\ref{Bcompstrans}) in equation~\eqref{reminder} gives
\begin{equation}
		\big\langle h_n\big|h_1\big\rangle
		=\frac{2\im}{\kappa^2\hbar}\int_{{M}\times\bC\bP^1\times\bC\bP^1}\hspace{-1.5cm}
		\rd\mu\ [\pi_n\,\rd\pi_n][\pi_1\,\rd\pi_1]\,B_{n\,\alpha}(F,\pi_n)\Lambda^{\alpha\gamma}(x,\pi_n)\,
		B_{1\,\beta}(F,\pi_1)\Lambda^\beta_{\ \gamma}(x,\pi_1)[\pi_n\,\pi_1]^3
\label{gammasqcomps}
\end{equation}
where ${M}\times\bC\bP^1\times\bC\bP^1$ is the fibrewise product of
$P(\mathbb{S}^+)$ with itself. The spinors $|\pi_1]$ and $|\pi_n]$
label to two copies of the $\bC\bP^1$ fibres.

\smallskip

This formula currently describes the scattering of two positive
helicity gravitons off a (fully non-linear) ASD background
spacetime. In order to obtain the BGK amplitudes, we must expand the
background spacetime ${M}$ around Minkowski space $\bM$.  In
principle, this can be done by iterating deformations of the twistor
space caused by adding in negative helicity momentum eigenstates, and
keeping track of the holomorphic degree-1 curves to construct the
function $F^\alpha(x,\pi)$ explicitly (see~\cite{Porter,Rosly} for a
discussion along these lines). In practice, constructing $F^\alpha$ in
this way is complicated, and the difficulties are compounded by having
to expand all the terms in~\eqref{gammasqcomps}.  Instead, motivated
by an analogous step at the same point in the Yang-Mills calculation
(equation~\eqref{twistorYM}), we seek a coordinate transformation of
the spin bundle $P(\bS^+)\to{M}$ that simplifies our task.

The desired coordinate transformation takes the form
\begin{equation} 
		(x^{\alpha\dot\alpha},\pi_{\dot\beta})\mapsto(y^{\alpha\dot\alpha}(x,\pi),\pi_{\dot\beta})
		\qquad\hbox{such that}\qquad
		\im y^{\alpha\dot\alpha}\pi_{\dot\alpha}=F^\alpha(x,\pi)\ ,
\label{coordtrans}
\end{equation}
and may be viewed as a $\pi$-dependent coordinate transformation of
${M}$. Equation~\eqref{coordtrans} replaces $F^\alpha$ by $\im
y^{\alpha\dot\alpha}\pi_{\dot\alpha}$, so that from the point of view
of the $(y,\pi)$ coordinates, we never need face the complicated
problem of constructing $F^{\alpha}(x,\pi)$ explicitly! The price to
be paid for this seemingly magical simplification is that generically,
the $\bC\bP^1$ fibres of $P(\bS^+)\to{M}$ do not coincide with those
of $P(\bS^+)\to\bM$ where here the $y$s are taken to be coordinates on
$\bM$; in other words, the $\bC\bP^1$s of constant $x$
(the twistor lines in $\cP\cT$) are not the same as the $\bC\bP^1$s of
constant $y$ (the twistor lines in $\bP\bT^\prime$).  There is some
freedom in the definition of $y^{\alpha\dot\alpha}$
in~\eqref{coordtrans}. One natural choice that fits the bill is
\begin{equation}
		 y^{\alpha\dot\alpha}(x,\pi)
		 =\im\frac{F^\alpha(x,\xi)\pi^{\dot\alpha}-F^\alpha(x,\pi)\xi^{\dot\alpha}}{[\xi\,\pi]}  
\label{ydef}
\end{equation}
where $\xi^{\dot\alpha}$ is an arbitrary constant spinor. Note that if the background is actually flat, then $F^\alpha=\im x^{\alpha\dot\alpha}\pi_{\dot\alpha}$ and we have simply
$y^{\alpha\dot\alpha}=x^{\alpha\dot\alpha}$. Also note that the
numerator vanishes at $|\pi]=|\xi]$, so the apparent singularity
when $[\xi\,\pi]=0$ is removable. Hence $y^{\alpha\dot\alpha}$ is smoothly (but not holomorphically) defined, and
$(y,\pi)$ are good coordinates on $P(\mathbb{S}^+)$, at least when the departure from flat spacetime is not too severe. Equation~\eqref{ydef} explicitly shows that the $\bC\bP^1$s of constant $x$ do not coincide with those of constant 
$y$, because $y$ varies as we move along a $\bC\bP^1$ fibre $L_x$.

We now pick a spacetime spin frame by requiring $\Lambda_\alpha^{\ \beta}(x,\xi)=\varepsilon_\alpha^{\ \beta}$. Then, using equation~\eqref{edef}, the Jacobian of the coordinate transformation~\eqref{ydef} with the spacetime tetrad
$\nabla^{(x)}_{\alpha\dot\alpha}$ is found to be
\begin{equation}
		\nabla^{(x)}_{\alpha\dot\alpha}y^{\beta\dot\beta} 
		 = \frac{1}{[\xi\,\pi]}\left(-\im\Lambda_\alpha^{\ \beta}(x,\pi)\pi_{\dot\alpha}\xi^{\dot\beta} 
		 - \varepsilon_\alpha^{\ \beta}\,\xi_{\dot\alpha}\pi^{\dot\beta}\right)\ .
\label{jacobian}
\end{equation}
This Jacobian has unit determinant because $\Lambda_\alpha^{\ \beta} \in SL(2,\bC)$, so 
$\rd\mu=\rd^4 y$ (mod $[\pi\,\rd\pi]$).  Furthermore, we see that 
\begin{equation}
		\pi^{\dot\alpha}\nabla^{(x)}_{\alpha\dot\alpha}
		=\pi^{\dot\alpha}\left(\nabla^{(x)}_{\alpha\dot\alpha}y^{\beta\dot\beta}\right)\frac{\del\ }{\del y^{\beta\dot\beta}}
		=\pi^{\dot\alpha}\frac{\del\ }{\del y^{\alpha\dot\alpha}}
\label{vectortrans}
\end{equation}
which will be used in what follows.

We are not quite ready to put this coordinate transformation to use,
because our expression~(\ref{gammasqcomps}) is written as an integral
over the fibrewise product of the spin bundle with itself, rather than
just as an integral over $P(\mathbb{S^+})$. Since~\eqref{coordtrans} does
not map the fibres of $P(\bS^+)\to{M}$ to the fibres of
$P(\bS^+)\to\bM$, if the coordinate transformation is given by say
$y(x, \pi_1)$, the $\pi_n$ integral in
equation~\eqref{gammasqcomps} will not hold $y$ constant for fixed $(x,\pi_1)$. 
We will deal with this by reformulating this
second fibre integral as an inverse of the $\delbar$-operator up the
$\bC\bP^1$ fibres of $P(\mathbb{S}^+)$ over ${M}$ and perturbing about
the the fibres of constant $y$.

We can understand $\delbar^{-1}$ as follows.  First recall that on a
single $\bC\bP^1$, any (0,1)-form is automatically $\delbar$-closed
for dimensional reasons. The cohomology groups
$H^{0,1}(\bC\bP^1,\cO(k))$ vanish for $k\geq-1$ and so any (0,1)-form
of homogeneity $k\geq-1$ is necessarily $\delbar$-exact. Thus, if
$\nu\in\Omega^{0,1}(\bP^1,\cO(k))$ with $k\geq -1$ then
$\delbar^{-1}\nu$ makes sense and is an element of
$\Omega^{0}(\bP^1,\cO(k))$. When $k\geq0$, $\delbar^{-1}\nu$ is not
uniquely defined because we can add to $\delbar^{-1}\nu$ a globally
holomorphic function $\rho$ of weight $k$, since
$\delbar\left(\delbar^{-1}\nu + \rho\right)=\nu$. However, there are
no global holomorphic functions of weight $-1$, so when $k=-1$,
$\delbar^{-1}\nu$ is unique. Explicitly, in terms of homogeneous
coordinates $\pi_{\dot\alpha}$ on the $\bC\bP^1$, one
takes\footnote{The numerical prefactor $1/2\pi\im$ of course involves
the ratio of a circle's area to its radius, rather than the
coordinate $\pi_{\dot\alpha}$.}
\begin{equation}
		\delbar^{-1}_{21} \nu:=\frac{1}{2\pi\im}\int_{\bC\bP^1}\frac{[\pi_1\,\rd\pi_1]}{[\pi_2\,\pi_1]}
\wedge \nu(\pi_1)\ , 
\end{equation}
which is indeed a 0-form of weight $-1$ in $\pi_2^{\dot\alpha}$. Taking $\delbar$
(with respect to $|\pi_2]$) of both sides shows that $\delbar\,
\delbar^{-1} = 1$, because the only $\pi_2$-dependence on the right
is from the homogeneous form 
$1/2\pi\im[\pi_2\,\pi_1]$ of the standard Cauchy kernel for
$\delbar^{-1}$.  (In affine coordinates $z$ on the Riemann sphere,
$\pi=(1,z)$ and $[\pi_2\,\pi_1]=z_1-z_2$.)

To exploit this in our situation, first use~\eqref{edef} \&~\eqref{Bcompstrans} to rewrite~\eqref{gammasqcomps} as
\begin{equation}
\big\langle h_n\big|h_1\big\rangle 
=\frac{2\im}{\kappa^2\hbar}\int_{{M}\times\bC\bP^1\times\bC\bP^1}\hspace{-1.5cm}
\rd\mu\ [\pi_n\,\rd\pi_n][\pi_1\,\rd\pi_1]\,
B_{n\,\alpha}(F,\pi_n)\Lambda^{\alpha\gamma}(x,\pi_n)\pi_n^{\dot\gamma}\tilde\nabla_{\gamma\dot\gamma}\,		\lrcorner\,\left(B_1(F,\pi_1)[\pi_n\,\pi_1]^2\right)\ .
\label{gammasqcomps2}
\end{equation}
Next, note that $p^*B_1\pi_1^{\dot\alpha}\pi_1^{\dot\beta}\pi_1^{\dot\gamma}$ has weight $-1$
in $|\pi_1]$ and is a $(0,1)$-form on (the second) $\bC\bP^1$ (valued also in $T^*{M} \otimes {\rm Sym}^3\,\bS^+$). We then define
\begin{equation}
		{}^{\phantom{x}}_x\delbar^{-1}_{n1}\left(B\,\pi^{\dot\alpha}\pi^{\dot\beta}\pi^{\dot\gamma}\right)
		:=\frac 1{2\pi\im}\int_{\bC\bP^1} \frac{[\pi_1\,\rd\pi_1]}{[\pi_n\,\pi_1]}\wedge
		B(F,\pi_1)\,\pi_1^{\dot\alpha}\pi_1^{\dot\beta}\pi_1^{\dot\gamma}
\label{delbar-1}
\end{equation}
where the presubscript $x$ emphasizes the fact that in this formula,
$\delbar$ involves the $(0,1)$-vector tangent to the $\bC\bP^1$ fibres  $q^{-1}(x)$.  
As above, this defines ${}_x\delbar^{-1}(B\,\pi^{\dot\alpha}\pi^{\dot\beta}\pi^{\dot\gamma})$
uniquely. Using this in equation~(\ref{gammasqcomps2}) allows us to
rewrite that equation as
\begin{equation}
	 	\big\langle h_n\big| h_1\big\rangle
		=-\frac{4\pi}{\kappa^2\hbar}\int_{P(\mathbb{S}^+)}\hspace{-0.5cm}\rd\mu\ [\pi_n\,\rd\pi_n]\  
		\Lambda^{\alpha\beta}B_{n\,\beta}(F,\pi_n) \pi_n^{\dot\alpha}\tilde\nabla_{\alpha\dot\alpha}\,\lrcorner\
		{}^{\phantom{x}}_x\delbar^{-1}_{n1}\left( B_1\,[\pi_n\,\pi_1]^3\right)\ ,
\label{gammasq}
\end{equation}
now interpreted as a (two-point) integral over the projective primed spin
bundle.

We can now use the coordinate transformation to simplify the
integral~(\ref{gammasq}).  Transforming to the $(y,\pi)$ coordinates using equations~\eqref{coordtrans} \&~\eqref{vectortrans} gives
\begin{equation}
		\big\langle h_n\big| h_1\big\rangle
		=-\frac{4\pi}{\kappa^2\hbar}\int_{P(\mathbb{S}^+)}\hspace{-0.4cm}\rd^4y\,[\pi_n\,\rd\pi_n]\ 
		B_n^\alpha(y,\pi_n)\pi_n^{\dot\alpha} \frac{\del\ }{\del y^{\alpha\dot\alpha}}\lrcorner\
		{}^{\phantom{x}}_x\delbar^{-1}_{n1}\,\!\left(B_1(y,\pi_1)[\pi_n\,\pi_1]^3\right)
\label{flatint} 
\end{equation}
now written as an integral on the spin bundle over flat spacetime. It
remains to reformulate the operator ${}_x\delbar^{-1}$, the inverse of
the $\delbar^{-1}$ operator on the $\bC\bP^1$s of constant $x$, in
terms of ${}_y\delbar^{-1}$ the inverse of the $\delbar$-operator on
the $\bC\bP^1$s of constant $y$.  These $\delbar$-operators are
essentially just the antiholomorphic tangent vector to the $\bC\bP^1$s
of constant $x$ or $y$ and the relationship between them follows by
the chain rule.  Using equations~(\ref{Fsolv}) \&~(\ref{ydef}) we find
\begin{equation}
		{}_x\delbar\,
		={}_y\delbar+ (\delbar y^{\alpha\dot\alpha})\frac{\del\ }{\del y^{\alpha\dot\alpha}} 
		= {}_y\delbar - \im\frac{p^*(V^\alpha)\xi^{\dot\alpha}}{[\xi\,\pi]}\frac{\del\ }{\del y^{\alpha\dot\alpha}}\ ,
\label{diffdelbar}
\end{equation}
where the extra term is the difference between an anti-holomorphic
vector field tangent to the fibres of $P(\bS^+)\to{M}$ and the anti-holomorphic vector field
tangent to the fibres of $P(\bS^+)\to\bM$.  Consequently, we see that\footnote{Equation~(\ref{diffdelbar}) is analogous to the Yang-Mills equation $\cA=-\delbar H\, H^{-1}$, while equation~(\ref{delbary}) is analogous to 
$H\delbar^{-1}H^{-1} = 1/(\delbar + \cA)$ used in appendix~\ref{sec:twym}.}  (somewhat
formally)
\begin{equation}
		\frac{1}{{}_x\delbar}= \frac{1}{{}_y\delbar +\cL_{\tilde V}}\ ,
\label{delbary}
\end{equation}
where the right hand side of this equation involves the $\delbar$-operator
along the $\bC\bP^1$ fibres in the $(y,\pi)$ coordinates, together with
the Lie derivative $\cL_{\tilde V}$ along the vector field
\begin{equation}
		\tilde{V}:=-\im\frac{p^*(V^\alpha)\xi^{\dot\alpha}}{[\xi\,\pi]}\frac{\del\ }{\del y^{\alpha\dot\alpha}}\ .
\label{spinbundV}
\end{equation}
(We will often abuse notation by not distinguishing $\tilde V$ from
its pushdown $p_*\tilde V = V$ to twistor space.) The Lie derivative
takes account of the fact that this operator acts on the form
$B_\alpha dy^{\alpha\dot\alpha}\pi_{\dot\alpha}$; both the components
$B_\alpha$ and basis forms $dy^{\alpha\dot\alpha}$ depend on $y$.

\begin{figure}
\begin{center}
\includegraphics[height=45mm]{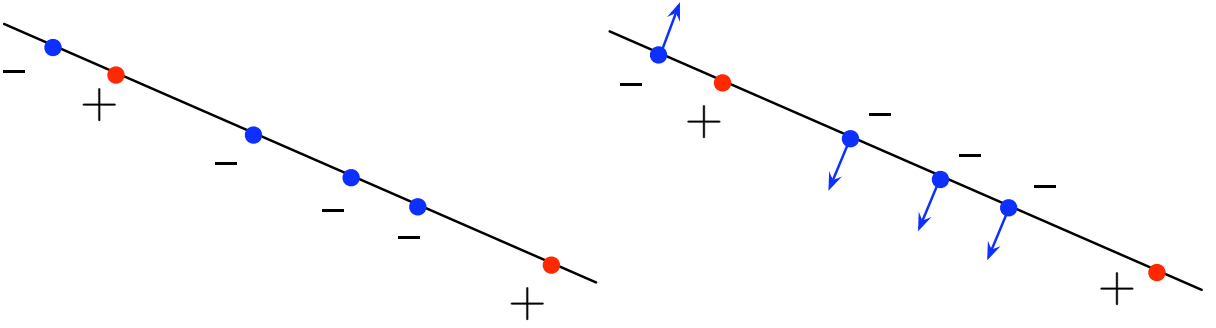}
\caption{{\it Yang-Mills ({\it l}) and gravitational ({\it r}) MHV
 amplitudes are supported on holomorphic lines in twistor space. For
 gravity, the negative helicity gravitons arise from insertions of
 normal vector fields, giving a perturbative description of the
 deformation of the line.}} 
\label{fig:support}
\end{center}
\end{figure}

The operator $(\delbar+\cL_V)^{-1}$ may be computed through its expansion
\begin{equation}
\frac{1}{\delbar+\cL_V} =\frac{1}{\delbar} - \frac{1}{\delbar}\cL_V\frac{1}{\delbar} +  
\frac{1}{\delbar}\cL_V\frac{1}{\delbar}\cL_V\frac{1}{\delbar} - \cdots
\label{dbartrans}
\end{equation}
where all the inverse $\delbar$-operators now imply an integral over
the $\bC\bP^1$s at constant $y^{\alpha\dot\alpha}$ (the holomorphic lines in $\bP\bT^\prime$). We have
\begin{equation}
		\big\langle h_n\big| h_1\big\rangle = 
		\sum_{n=2}^\infty(-)^{n+1}\frac{4\pi}{\kappa^2\hbar}
		 \int \rd^4y\,[\pi_n\,\rd\pi_n]\ B^\alpha_n\pi_n^{\dot\alpha}\frac{\del\ }{\del y^{\alpha\dot\alpha}}\,\lrcorner
		 \left(\frac{1}{\delbar}\cL_{V_{n-1}}\frac{1}{\delbar}\cdots\frac{1}{\delbar}\cL_{V_2}\frac{1}{\delbar}
		 B_1[\pi_n,\pi_1]^3\right)\ .
\end{equation}
The inverse $\delbar$-operators always act on sections of
$\Omega^{0,1}(\bC\bP^1,\cO(-1)\otimes T^*M)$ and so are canonically
defined as in equation~(\ref{delbar-1}), although here it is $y$ rather
than $x$ that is being held constant. Because the vector fields
$\tilde V$ point in the $y$-direction, the Lie derivatives may be
brought inside all the $\bC\bP^1$ integrals, effectively commuting
with the inverse $\delbar$-operators. So the $n^{\rm th}$-order term
in the expansion is   
\begin{equation}
		\cM^{(n)}_{\rm twistor} := \frac{\im^n}{(2\pi)^{n-2}}\frac{4\pi}{\kappa^2\hbar}
		\int \rd^4y\,\prod_{i=1}^n\frac{[\pi_i\,\rd\pi_i]}{[\pi_{i+1}\,\pi_i]}\ B^\alpha_n\pi_n^{\dot\alpha}
		\frac{\del\ }{\del y^{\alpha\dot\alpha}}\,\lrcorner\left(\cL_{V_{n-1}}\cdots\cL_{V_2}
		B_1[\pi_n,\pi_1]^4\right)
\label{gravexpand}
\end{equation}
where we have compensated for the fact that the integration measure
$\prod_{i=1}^n[\pi_i\,\rd\pi_i]/[\pi_{i+1}\,\pi_i]$ includes an extra
factor of $1/[\pi_n\,\pi_1]$ by increasing the power of
$[\pi_n\,\pi_1]$ in the numerator. In this expression the $n$-point
amplitude comes from an integral over the space of lines twistor
space, with $n$ insertions on the line, each of whose insertion point
is integrated over. This is exactly the same picture as described in
the appendix for
Yang-Mills. For gravity, the $n-2$ vector fields differentiate the
wavefunctions (as we will see explicitly later), leading to what is
sometimes called `derivative of a $\delta$-function' support (see
figure~\ref{fig:support}).

\smallskip

Our final task is to evaluate this expression when the external states
are each the plane waves of \eqref{planewave}\eqref{Bplanewave}.
From~(\ref{Vdef}), the associated twistor space vector fields are
$V(Z)=\kappa\,\bar\delta_{(1)}([\pi\, k])\,\e^{\langle\omega\, \tilde
k\rangle}\,\tilde{k}^\alpha\del/\del\omega^\alpha$, so the vector
fields on $P(\mathbb{S}^+)$ become
\begin{equation}
		V(y,\pi)=-\im\kappa\,\bar\delta_{(1)}([\pi\, k])\,\exp\left(\im p\cdot y\right)
		\frac{\tilde k^\alpha\xi^{\dot\alpha}}{[\xi\,\pi]}\frac{\del\ }{\del y^{\alpha\dot\alpha}}
\label{Vcomps}
\end{equation}
using equation~(\ref{spinbundV}). 
Pulling the plane wave formula \eqref{Bplanewave} for $B$ back to $P(\bS^+)$
gives
\begin{equation}
		B = \im\kappa\,\frac{\langle\tilde\beta|dy|\pi]}{\langle\tilde\beta\,\tilde k\rangle}
		\bar\delta_{(-5)}([\pi\,k])\,\exp\left(\im p\cdot y\right)
\label{Bcomps}
\end{equation}
in the $(y,\pi)$ coordinates.

To evaluate~\eqref{gravexpand}, use the Cartan formula
$\cL_{V}=V\lrcorner\,\rd+\rd V\lrcorner$ to replace $\cL_{V_{n-1}}$. The
second term in Cartan's formula leads to a contribution
\begin{equation}
		B_n^\alpha\pi_n^{\dot\alpha}\frac{\del\ }{\del y^{\alpha\dot\alpha}}\,\lrcorner\, 
		\rd\left(V_{n-1}\,\lrcorner\,\cL_{V_{n-2}}\cdots\cL_{V_2}B_1\right)
		=B_n^\alpha\pi_n^{\dot\alpha}\frac{\del\ }{\del y^{\alpha\dot\alpha}}
		\left(V_{n-1}\,\lrcorner\,\cL_{V_{n-2}}\cdots\cL_{V_2}B_1\right)
\label{totalderiv}
\end{equation}
to the integrand of~\eqref{gravexpand}. On the right hand side, 
$B_n^\alpha\pi_n^{\dot\alpha}\del/\del y^{\alpha\dot\alpha}$ simply differentiates the scalar $V_{n-1}\,\lrcorner\,\cL_{V_{n-2}}\cdots\cL_{V_2}B_1$. Because $B_n$ is pulled back to $P(\mathbb{S}^+)$ from twistor space, it depends on $y$ only through $y^{\alpha\dot\alpha}\pi_{\dot\alpha}$, so $B_n^\alpha$ may be brought inside the
$\pi_n^{\dot\alpha}\del/\del y^{\alpha\dot\alpha}$ derivative. Hence~\eqref{totalderiv} is a total derivative and may be discarded. Now, using the fact that $[\rd,\cL_V]=0$ for any vector field $V$, the remaining terms involve
\begin{multline}
		B_n^\alpha\pi_n^{\dot\alpha}
		\frac{\del\ }{\del y^{\alpha\dot\alpha}}\,\lrcorner\,V_{n-1}\,\lrcorner\,\left(\rd\cL_{V_{n-2}}\cdots\cL_{V_2}B_1
		\right)\\
		=B_n^\alpha\pi_n^{\dot\alpha}
		\frac{\del\ }{\del y^{\alpha\dot\alpha}}\,\lrcorner\,V_{n-1}\,\lrcorner\left(\cL_{V_{n-2}}\cdots\cL_{V_2} 
		\left(\frac{\tilde h_1}{2} \rd y^{\gamma\dot\gamma}\wedge \rd y_\gamma^{\ \dot\delta}
		\pi_{1\dot\gamma}\pi_{1\dot\delta}\right)\right)
\label{gravexpand2}
\end{multline}
where we have used $\rd B=\left(\tilde h/2\right)\rd y^{\alpha\dot\alpha}\wedge \rd y_{\alpha}^{\ \dot\beta}\pi_{\dot\alpha}\pi_{\dot\beta}$, which again follows because $B$ is pulled back from a field on twistor space.

The key simplification that allows us to evaluate~\eqref{gravexpand2} comes from making the gauge choice $|\xi]=|n]$, where $|n]$ is the dotted momentum spinor of  the positive helicity graviton represented by $B_n$. With this choice, the two-form
$\cL_{V_{n-2}}\cdots\cL_{V_2}\left(\tilde h_1/2\, \rd y^{\gamma\dot\gamma}\wedge  \rd y_\gamma^{\ \dot\delta}
\pi_{1\dot\gamma}\pi_{1\dot\delta}\right)$ is contracted into the bi-vector $B_n^{\alpha}\pi_n^{\dot\alpha}V_{n-1}^{\beta}
n^{\dot\beta}\left(\del/\del y^{\alpha\dot\alpha}\wedge\del/\del y^{\beta\dot\beta}\right)$. But the momentum eigenstate $B_n$ has support only when $|\pi_n]=|n]$, so this bi-vector is purely self-dual:
\begin{equation}
		B^\alpha\pi_n^{\dot\alpha}\frac{V_{n-1}^\beta n^{\dot\beta}}{[n\,\pi_{n-1}]}
		\frac{\del }{\del y^{\alpha\dot\alpha}}\wedge\frac{\del\ }{\del y^{\beta\dot\beta}}
		=\frac{1}{2}\frac{\langle B_n\, V_{n-1}\rangle}{[n\,\pi_{n-1}]}\,\pi_n^{\dot\alpha}n^{\dot\beta}
		\frac{\del\ }{\del y^{\beta\dot\beta}}\wedge\frac{\del\ }{\del y_\beta^{\ \dot\alpha}}\ .
\label{sdcontract}
\end{equation}
It is straightforward to check that because the vectors $V_i$ are Hamiltonian, with our gauge choice, the 
bi-vector $\pi_n^{\dot\alpha}n^{\dot\beta}\,\del/\del y^{\alpha\dot\alpha}\wedge\del/\del y_\alpha^{\ \dot\beta}$ commutes with all the remaining Lie derivatives. Therefore, we may immediately contract this bivector with $\rd y^{\gamma\dot\gamma}\wedge\rd y_\gamma^{\ \dot\delta}\,\pi_{1\dot\gamma}\pi_{1\dot\delta}$ to obtain 
\begin{equation}
		\cM^{(n)}_{\rm twistor}=\frac{\im^n}{(2\pi)^{n-2}}\frac{2\pi}{\kappa^2\hbar}
		\int \rd^4y\,\prod_{i=1}^n\frac{[\pi_i\,\rd\pi_i]}{[\pi_{i+1}\,\pi_i]}\ \frac{\langle B_n\,V_{n-1}\rangle}{[n\,\pi_{n-1}]}
		V_{n-2}\cdots V_2(\tilde h_1)[\pi_n\,\pi_1]^5[n\,\pi_1]^5
\label{amps}
\end{equation}
where the remaining vector fields $V_2$ to $V_{n-2}$ act simply by differentiating everything to their right.

To take account of the possible orderings of the external states, we insert
\begin{equation}
		V_m =\kappa\sum_{i=2}^{n-1}\epsilon_i\,
				\bar\delta_{(1)}([\pi\, i])\,\e^{\im p_i\cdot y}\,
				\frac{\tilde{i}^\alpha n^{\dot\alpha}}{[n\,\pi_m]}\frac{\del\ }{\del y^{\alpha\dot\alpha}}
\end{equation}
for each vector field $V_m$ at $(y^{\alpha\dot\alpha},\pi_m^{\dot\beta})$, where the $\epsilon_i$ are expansion parameters labelling the physical external states. (We use the shorthand $p_i^{\alpha\dot\alpha}=\tilde i^{\alpha} i^{\dot\alpha}$.) Extracting the coefficient of $\prod_{i=2}^{n-1}\epsilon_{i}$ and using the $\bar\delta$-functions to integrate over the $n$ insertion points gives the $n$-particle MHV amplitude as
\begin{multline}
		\cM^{(n)}_{\rm twistor}
		= \frac{\kappa^{n-2}}{\hbar}\delta^{(4)}\left(\sum_{i=1}^n p_i\right)[1\,n]^8\ \times\\
		\left\{\frac{\langle\tilde\beta\,n-1\rangle}{\langle\tilde\beta\,n\rangle[n-1\,n][n\,1]^2}\,\frac{1}{C(n)}\,
		\prod_{k=2}^{n-2}\frac{\langle k|p_{k-1}+p_{k-2}+\cdots +p_{2}+p_1|n]}{[n\,k]} 
		\ +\ {\rm P}_{(2,\ldots,n-1)}\right\}\ ,
\label{twistamps}
\end{multline}
where ${\rm P}_{(2,\dots,n-1)}$ is a sum over permutations of the vector fields. Consider the first (displayed) permutation. This is the same as the first term in $\cM^{(n)}$ in equation~(\ref{amplitudes}), except for a factor
\begin{equation}
		\frac{\langle\tilde\beta\,n-1\rangle}{\langle\tilde\beta\,n\rangle[n\,1]}\times[1\,n-1]
		= -\frac{\langle\tilde\beta| p_{n-1}|1]}{\langle\tilde\beta| p_n|1]}\ .
\end{equation}
This factor is independent of $2,\ldots,n-2$, permuting the first term over gravitons $2$ to $n-2$ will yield the same
factor times the corresponding permutation of~(\ref{amplitudes}). Therefore we have
\begin{equation}
		\cM^{(n)}_{\rm twistor}
		= -\frac{\langle\tilde\beta| p_{n-1}|1]}{\langle\tilde\beta| p_n|1]}\cM^{(n)}+\hbox{other perms}\ .
\label{relation}
\end{equation}
The remaining permutations in~(\ref{relation}) involve exchanging graviton $n-1$ with each of gravitons $2$ to $n-2$. 
But since $\cM^{(n)}$ is equal to the standard BGK amplitude (as proved in appendix~\ref{sec:BGK}), we know ({\it e.g.} from Ward identities~\cite{BDDPR}) that it is in fact symmetric under exchange of any two like-helicity gravitons. Hence each term is proportional to $\cM^{(n)}$ and we are left with an overall factor
\begin{equation}
		-\sum_{i=2}^{n-1} \frac{\langle\tilde\beta|p_i|1]}{\langle\tilde\beta|p_n|1]} 
		= -\frac{\langle\tilde\beta|p_2+p_3+\cdots+p_{n-1}|1]}{\langle\tilde\beta|p_n|1]}=1.
\end{equation}
Thus we have shown that~(\ref{twistamps}) is really independent of
$\tilde\beta$, and that $\cM^{(n)}_{\rm twistor}=\cM^{(n)}$. 

It is remarkable that the infinite series of $n$-particle MHV
amplitudes may be constructed by expanding the square of the self-dual spin connection on
an anti self-dual spacetime \eqref{reminder}.


\section{A Twistor Action for MHV Diagrams in Gravity}
\label{sec:twistact}

According to the MHV {\it diagram} formalism, initiated
in~\cite{CSW} for Yang-Mills and~\cite{BBDIPR,BB-BD,Nasti} for
gravity, one can recover the full perturbation theory by continuing
the MHV amplitudes off-shell and connecting them together using
propagators connecting positive and negative helicity
lines\footnote{At the quantum level, this program works as stated only
for supersymmetric theories~\cite{Trees2Loops}.}. The MHV diagram
formalism was first developed in the context of the `disconnected prescription' of
twistor-string theory~\cite{CSW}, but soon after it was realized that one could
also construct actions whose Feynman diagrams generate the Yang-Mills
MHV diagram formalism~\cite{Mason,BMS,BMS2,Boels,Mansfield,Ettle,GorskyRosly}. We
now give a twistor action whose perturbation theory generates the MHV
diagram formalism for gravity.

In section~\ref{sec:nonlineargrav}, ASD spacetimes were reformulated
in terms of deformed twistor spaces by the nonlinear graviton
construction~\cite{Penrose}. The field equation on twistor space
is the vanishing of the Nijenhuis tensor
\begin{equation}
N = I^{IJ}\del_J\left(\delbar h + \frac{1}{2}\left\{h,h\right\}\right)
\label{Nijenhuis2}
\end{equation}
so that the almost complex structure $\delbar + I(\rd
h,\,\cdot\,)$ is integrable and $\cP\cT$ is a complex threefold,
obtained as a  deformation of $\bP\bT^\prime$ (see the discussion around
equation~\eqref{delbardef}).  In~\cite{MW}, a local twistor action
whose field equations include the condition $N=0$ was constructed. The
action is written in terms of a field\footnote{We abuse notation
by not distinguishing the (0,1)-forms $h$, $\tilde h$ from their
cohomology classes.} $h\in\Omega^{0,1}(\bP\bT^\prime,\cO(2))$ and
$\tilde h\in\Omega^{0,1}(\bP\bT^\prime,\cO(-6))$, although we 
also here use 
$B\in\Omega^{1,1}(\bP\bT^\prime,\cO(-4))$. It takes a `BF'-like form
\begin{equation}
S = \int_{\bP\bT^\prime} \Omega\wedge I^{IJ}B_I\del_J\left(\delbar h
+\frac{1}{2}\left\{h,h\right\}\right) = -
\int_{\bP\bT^\prime}\Omega\wedge \tilde h\left(\delbar h +
\frac{1}{2}\left\{h,h\right\}\right) 
\label{MW}
\end{equation}
where $\Omega=\epsilon_{IJKL}Z^I\rd Z^J\wedge\rd Z^K\wedge\rd Z^L/4!$
is the canonical holomorphic 3-form of weight $+4$, $I^{IJ}$ is the
Poisson structure introduced in equation~\eqref{infinity} and
$\{\,\cdot\,,\,\cdot\,\}$ its associated Poisson bracket. Note that
this Poisson bracket has weight $-2$ so that the action is
well-defined on the projective space.  In the first version, $B$ plays
the r\^ole of a Lagrange multiplier ensuring the vanishing of the
Nijenhuis tensor. The second form follows upon integration by parts.

In the second form, the field equations of this action are
\begin{equation}
\delbar h + \frac12 \{h,h\} = 0\qquad\hbox{and}\qquad \delbar_h\tilde
h = 0\, , \quad \mbox{ where } \quad \delbar_h f:=\delbar f +
\{h,f\}\, .
\end{equation}
We also have the gauge freedom $h \to h + \delbar_h\chi$, $\tilde
h\to\tilde h +\delbar_h\tilde\chi$.  In the linearized theory, these
imply that on-shell, $h$ and $\tilde h$ are representatives of the
cohomology classes used to described linearized gravitons of
helicities $\pm2$ in the Penrose transform, as reviewed in
section~\ref{sec:lineartwistorgrav}. The gauge freedom may be fixed by
using\footnote{The Newman gauge of section~\ref{sec:nonlineargrav}
implies CSW gauge, but also enforces other conditions appropriate
only when the fields are on-shell.}  `CSW gauge'~\cite{CSW,BMS}:
choose an antiholomorphic vector field $\bar\eta$ tangent to the
fibres of $\bP\bT^\prime\to\bC\bP^1$ and impose the axial gauge
condition that $\bar\eta\,\lrcorner\, h=0$ and
$\bar\eta\,\lrcorner\,\tilde h =0$. 
Imposing this gauge in~\eqref{MW}, the cubic vertex
vanishes and one is left with an off-diagonal kinetic term and a
linear theory.

The other main ingredient in the MHV diagram formulation is the
infinite set of MHV vertices: off-shell continuations of the MHV
amplitudes. Using coordinates $(y,\pi)$ for the spin bundle
$P(\bS^+)\to\bM$ over Minkowski space, it follows from the previous
section that in the twistor formulation these vertices arise from the
expansion of
\begin{equation}
\int_{P(\bS^+)}\hspace{-0.3cm}\rd^4y\wedge[\pi_n\,\rd\pi_n]
\wedge B^\alpha(y,\pi_n)\pi_n^{\dot\alpha}\frac{\del\ }{\del y^{\alpha\dot\alpha}}
\,\lrcorner\,\left(\frac{1}{\delbar+\cL_{\tilde V}} B(y,\pi_1)
[\pi_n\,\pi_1]^3\right) 
\label{MHVvertices}
\end{equation}
where we interpret $B$ as the pullback to $P(\bS^+)$ 
of an arbitrary element of
$\Omega^{1,1}(\bP\bT^\prime,\cO(-4))$ ({\it i.e.}, not necessarily
obeying $\delbar B=0$). Likewise, $\tilde V$ is here interpreted as
in \eqref{spinbundV}:
\be\label{spinbundV2}
\tilde V= -\im \frac{p^* (V^\alpha) \xi^{\dot\alpha}}{[\xi\, \pi]}\frac
\del{\del y^{\alpha\dot\alpha}}\quad \mbox{ where } \quad
V_{\alpha}=\frac{\del h}{\del \omega^\alpha}\, .
\ee
The inverse operator $1/(\delbar+\cL_{\tilde V})$ is again understood
through its infinite series expansion~\eqref{dbartrans} leading to an
infinite sequence of MHV 
vertices.  These only involve the components of the (0,1)-forms $B$ and $h$
that are tangent to the $\bC\bP^1$ base of $\bP\bT^\prime\to\bC\bP^1$.

The choice of the vector field $\bar\eta$ corresponds to the choice of
spinor used by~\cite{CSW}.  As described for Yang-Mills
in~\cite{BMS2}, it enters into the definition of the propagator which
gives the CSW rule for extending the MHV amplitudes off-shell.  The
discussion in ~\cite{BMS2} applies here directly with just a shift in
homogeneities.  Therefore, treating $h$ and $B$ as the fundamental
fields, in CSW gauge, the Feynman diagrams of the action
\begin{multline}
S[B,h] = \int_{\bP\bT^\prime} \Omega\wedge
I^{IJ}B_I\del_J\left(\delbar h + \frac{1}{2}\{h,h\}\right)\\
+\int_{P(\bS^+)}\hspace{-0.3cm}\rd^4y\wedge[\pi_n\,\rd\pi_n] \wedge
B^\alpha(y,\pi_n)\pi_n^{\dot\alpha}\frac{\del\ }{\del
  y^{\alpha\dot\alpha}}
\,\lrcorner\,\left(\frac{1}{\delbar+\cL_{\tilde V}} B(y,\pi_1)
  [\pi_n\,\pi_1]^3\right)
\label{twistgravact}
\end{multline}
reproduces the MHV diagram formulation of gravity.

\section{Supergravity}
\label{sec:sugra}
Supertwistor space $\bP\bT^\prime_{[\cN]}$ is the projectivisation of
$\bC^{4|\cN}$ where we have adjoined $\cN$ anticommuting homogeneity
degree 1 coordinates $\psi^A$, $A=1,\ldots ,\cN$.  In Penrose
conventions, the space of holomorphic lines in $\bP\bT^\prime_{[\cN]}$
is anti-chiral superspace $\bM_{[\cN]}$ with coordinates
$(x^{\alpha\dot\alpha}, \tilde\theta^{A\dot\alpha})$, where
$\tilde\theta^{A\dot\alpha}$ are anti-commuting. The flat space
incidence relation~\eqref{incidef} is augmented to
\begin{equation}
		\omega^\alpha=\im x^{\alpha\dot\alpha}\pi_{\dot\alpha}\qquad
		\psi^A=\tilde\theta^{A\dot\alpha}\pi_{\dot\alpha}\ .
\end{equation}

The linear Penrose transform of section~\ref{sec:lineartwistorgrav} extends~\cite{Ferber} to one between cohomology classes on $\bP\bT^\prime_{[\cN]}$ and superfields on $\bM_{[\cN]}$.  In particular, $h$ naturally extends to an (on-shell)
superfield\footnote{$\bP\bT^\prime_{[\cN]}$ is a split supermanifold, whose cohomology is generated by that of the base.} $\cH\in H^1(\bP\bT^\prime_{[\cN]},\cO(2))$ that is holomorphic in $\psi^i$. That is, $\cH$ has component expansion
\begin{equation}
		\cH(Z,\psi) = h(Z) + \psi^A\lambda_A(Z) + \cdots + (\psi^1\psi^2\cdots\psi^\cN) \phi(Z)
\label{superexpandH}
\end{equation}
where the coefficient of $(\psi)^k$ may represented by a (0,1)-form on
the standard twistor space $\bP\bT^\prime$ and has homogeneity $2-k$.
$\cH$ generates Poisson deformations of the complex structure of the
twistor superspace through its associated Hamiltonian vector
superfield $I(\rd\cH,\,\cdot\,)$~\cite{MW}.  As a superfield it
represents on-shell spacetime fields of helicities
$-2,-\frac{3}{2},\ldots,-2+\frac{\cN}{2}$. When $\cN<8$, the conjugate
graviton supermultiplet is represented by a twistor superfield
$\tilde\cH\in H^1(\bP\bT^\prime_{[\cN]},\cO(\cN-6))$. As in the
non-supersymmetric case, $\tilde\cH$ may equivalently be represented by a
superfield $\cB\in H^1(\bP\bT_{[\cN]},\Omega^1(\cN-4))$, modulo the
gauge equivalence $\cB\to\cB+\rd m(Z,\psi)+n(Z,\psi)[\pi\,\rd\pi]$.

A particularly interesting case is $\cN=4$, for which twistor space is
a Calabi-Yau supermanifold, {\it i.e.}, it admits a global holomorphic
volume (integral) form.
The Calabi-Yau property
singles out $\cN=4$ twistor space as a natural target for a string
theory~\cite{Witten}. When $\cN=4$, $\cB(Z,\psi)$ has homogeneity zero
and there is a natural extension of the action~\eqref{twistgravact}:
\begin{multline}
		S_{\cN=4}=\int_{\bP\bT^\prime_{[4]}}\hspace{-0.3cm}\rd^4\psi\,\Omega\wedge I^{IJ}\cB_I\del_J
		\left(\delbar\cH+\frac{1}{2}\{\cH,\cH\}\right)\\
		+\int_{P(\bS^+_{[4]})}\hspace{-0.3cm}\rd^{4|8}x\wedge[\pi_n\,\rd\pi_n]\wedge
		\frac{\cB_n^{\alpha}\xi^{\dot\alpha}}{[\xi\,\pi_n]}\frac{\del\ }{\del x^{\alpha\dot\alpha}}
		\,\lrcorner\,\left(\frac{1}{\delbar+\cL_{\cV}}\,\cB\right)\, .
\label{N=4act}
\end{multline}
where in the second term, $\cB$ is pulled back to the superspace spin
bundle $P(\bS^+_{[4]})$ and $\cV$ is the vector field on
$P(\bS^+_{[4]})$ defined by \eqref{spinbundV} (or \eqref{spinbundV2})
with $V$ replaced by $\cV$.  In this $\cN=4$ formula, the inverse
$\delbar$-operators act on (0,1)-forms of vanishing weight. Although
$\delbar^{-1}$ is not obstructed on such forms, it is ambiguous. The
freedom can be fixed by adding a constant so that
$\delbar^{-1}_{21}\cB$ vanishes when $|\pi_2]=|\xi]$. With this
choice,
\begin{equation}
		\delbar^{-1}_{21}\cB(x,\tilde\theta,\pi_2):= \frac{1}{2\pi\im}\int_{\bC\bP^1}\frac{[\pi_1\,\rd\pi_1]}{[\pi_2\,\pi_1]}
		\frac{[\xi\,\pi_2]}{[\xi\,\pi_1]}\ \cB(x,\tilde\theta,\pi_1)
\end{equation}
which has homogeneity zero in $|\pi_2]$ and satisfies $\delbar\delbar^{-1}\cB = \cB$; the integrand in the second term 
of~\eqref{N=4act} then has vanishing weight in each $\bC\bP^1$ and is thus well-defined. It is easy to check that the truncation of~\eqref{N=4act} to $\cN=0$ reproduces~\eqref{twistgravact}. As in~\cite{Nair1}, when $\cB$ and 
$\cV$ are on-shell with respect to the local $\cN=4$ twistor action and are taken to be the twistor momentum eigenstates
\begin{eqnarray}
		\cV(Z,\psi) &=& \kappa\,\bar\delta_{(1)}([\pi\, k])\,\exp\left(\langle\omega\tilde k\rangle + \psi^A\zeta_A\right)
				\ \tilde k^\alpha\frac{\del\ }{\del \omega^\alpha} \label{superV}\\
		\cB(Z,\psi) &=& \kappa\,\bar\delta_{(-5)}([\pi\,k])\,\exp\left(\langle\omega\tilde k\rangle + \psi^A\zeta_A\right)
				\frac{\langle\tilde\beta\,\rd\omega\rangle}{\langle\tilde\beta\,\tilde k\rangle}\ ,
				\label{superB}
\end{eqnarray}
then the coefficients of the external Grassmann parameters $\zeta_A$ in an expansion of the non-local term give 
the MHV amplitudes for arbitrary external members of the $\cN=4$ supermultiplet.

Although $\cN=4$ twistor supersymmetry seems natural in twistor-string theory, $\cN=8$ supergravity is usually thought of as more fundamental. The $\cN=8$ graviton supermultiplet is CPT self-conjugate, and this fact has recently been argued to underlie many surprisingly simplifications in the S-matrix~\cite{FreddyNima}. Thus, on twistor space, the complete multiplet is represented by a single superfield
\begin{equation}
		\cH(Z,\psi) = h(Z) + \psi^A\lambda_A(Z)+\cdots+(\psi)^8\,\tilde h(Z)
\label{N=8multiplet}
\end{equation}
In the case that the external states are on-shell momentum eigenstates, represented on twistor space by the Newman gauge expression~\eqref{superV}, the MHV scattering of arbitrary members of the $\cN=8$ multiplet is described by the formula
\begin{equation}
		\cM^{(n)}_{\cN=8} = \int_{P(\bS^+_{[8]})} \hspace{-0.3cm}\rd^{4|16}x\ 
		\prod_{i=1}^n\frac{[\pi_i\,\rd\pi_i]}{[\pi_{i+1}\,\pi_i]} 
		\frac{\cH_n}{[\pi_1\,\pi_n]}	\frac{\cH_{n-1}}{[\pi_n\,\pi_{n-1}]}\,\cV_{n-2}\cdots
		\cV_2\left(\frac{\cH_1}{[\pi_{n-1}\,\pi_1]}\right)\ .
\label{superamps}
\end{equation}
Unlike the previous formul\ae~\eqref{gravexpand} \&~\eqref{N=4act},
this expression singles out {\it three} of the external fields,
representing them in terms of the Hamiltonian function $\cH$ rather
than the vector field $\cV$. This is closely related to the formula obtained
by Nair in~\cite{Nair3}. It is easy to check that~\eqref{superamps}
reproduces the BGK amplitudes for external gravitons, and satisfies
the supersymmetric recursion relations of~\cite{BEF}.


\section{Conclusions and future directions}

A perspective of this paper has been that the MHV vertices provide a
bridge between perturbative treatments of gravity and the fully
nonlinear, non-perturbative structure that is such a key part of
General Relativity. When we are on-shell with respect to the chiral
action~\eqref{MW} (or the chiral limit of the Plebanski action), we
may take advantage of the integrability of the anti self-dual Einstein
equations to interpret the infinite sum of MHV amplitudes as simply
the square of a linearized fluctuation $\gamma$ of the self-dual spin
connection $\Gamma $ on the ASD background. The techniques of this
paper, both for Yang-Mills and gravity indicate that it is possible to
develop a background field formalism on fully nonlinear asd
backgrounds within which explicit computations are tractable and
generate amplitudes for processes with an arbitrary number of negative
helicity legs.  This programme would allow one to incorporate the
integrability of the anti self-dual Einstein equations into the study
of perturbation theory in such a way as to bridge the gap between
perturbative and non-perturbative treatments of gravity.

The status of the MHV diagram formulation for gravity is currently
less clear than that for Yang-Mills, although it has now been verified
for up to 11 external particles~\cite{BEF}.  At this stage there
is no reason to doubt that the MHV picture for gravity should be
successful, at least classically. The validity of our twistor
action~\eqref{twistgravact} for gravity currently depends on that of
the MHV formalism whereas, in the case of Yang-Mills, the twistor
action of~\cite{Mason,BMS} and reviewed in appendix~\ref{sec:ym}
provides an independent non-perturbative derivation of the MHV
formalism~\cite{BMS2}.  A future goal is to construct a twistor
action for gravity that works in the same way---for this it will be
necessary to build a formalism in which the background is {\it
off-shell} and $\cP\cT$ possesses only an {\it almost} complex
structure.  A search for a spacetime MHV Lagrangian for gravity has
been initiated in~\cite{ATh}, following the path of~\cite{Mansfield}
in Yang-Mills.

$\cN=4$ supergravity is not unique, and~\eqref{N=4act} is not the
unique $\cN=4$ completion of the non-supersymmetric action. Firstly,
the Poisson structure $I$ may also point along the fermionic
directions, and in~\cite{MW,Martin} this was shown to be responsible
for gauged supergravities in the self-dual sector. Secondly,
unlike the $\cN=4$ completion of the MHV amplitudes in Yang-Mills,
there seems to be no compelling reason that the nonlocal term
in~\eqref{N=4act} should be only {\it quadratic} in $\cB$. It would be
interesting to know if additional terms are required in the case of
gauged supergravity.

A key motivation for much of the work here is to reverse engineer a
twistor-string theory for gravity.  The Lie derivatives and inverse
$\delbar$-operators in the second term in the action~\eqref{N=4act}
are suggestive of a worldsheet OPE interpretation, and it would be
fascinating to see if this term (taken on-shell) can arise as an
instanton contribution in some form of twistor-string theory.  In
particular, $h$ and $B$ enter just as they do in the vertex operators
of~\cite{AHM}.

\subsection*{Acknowledgements}
We would like to thank Mohab Abou-Zeid, Paolo
Benincasa, Rutger Boels, Freddy Cachazo, Henriette Elvang, Dan Freedman and Chris Hull
for useful discussions. LM is partially supported by the EU through
the FP6 Marie Curie RTN {\it ENIGMA} (contract number
MRTN--CT--2004--5652) and through the ESF MISGAM network. This work
was financed by EPSRC grant number EP/F016654,\\
http://gow.epsrc.ac.uk/ViewGrant.aspx?GrantRef=EP/F016654/1.

\appendix

\section{Simplifying the BGK Amplitudes}
\label{sec:BGK}

In this appendix we will show analytically that the Berends, Giele \&
Kuijf~\cite{BGK} form of the graviton MHV amplitude agrees with the
simplified expression~(\ref{amplitudes}) used in the text. Similar manipulations have
been performed in~\cite{Nair3,BBST,EF}; our version of the amplitude
is nearest to one given implicitly in~\cite{Nair3}, although we
believe the detailed form is new.

Berends, Giele \& Kuijf give the MHV amplitude
\begin{equation}
		\cM_{\rm BGK} = \frac{\kappa^{n-2}}{\hbar}\, \delta^{(4)}\left(\sum p\right)\,M\ ,
\end{equation}
where for $n\geq5$
\begin{equation}
		M(1^+,2^-,3^-,\ldots, n-1^-,n^+)= [1n]^8\ 
		\left\{\frac{\langle 12 \rangle\langle
                n-2\,n-1\rangle}{[1\,n-1]}\frac{F}{N(n)} 
		\prod_{i=1}^{n-3}\prod_{j=i+2}^{n-1}[ij] + {\rm
              P}_{(2,\ldots,n-2)}\right\} 
\label{BGK}
\end{equation} 
with $N(n):=\prod_{i<j}\ [ij]$ and where
\begin{equation}
		F := \prod_{k=3}^{n-3}\ \langle k|p_{k+1}+p_{k+2}+ \cdots +p_{n-1}|n]
\label{Fdef}
\end{equation}
when $n\geq6$ and $F=1$ when $n=5$. In~(\ref{BGK}), the symbol ${\rm P}_{2,\ldots,n-2}$ denotes a sum over all 
permutations of gravitons $2$ to $n-2$. 

We begin by writing
\begin{equation}
\begin{aligned}
		\frac{\langle 12 \rangle\langle n-2\,n-1\rangle}{[1\,n-1]}
		&=\frac{\langle 21 \rangle[1n]\ \langle n-2\,n-1\rangle[n-1\,n]}{[1\,n-1][n-1\,n][n1]}\\
		&=-\frac{\langle 2|p_3+p_4+\cdots+p_{n-1}|n]\langle n-2|p_{n-1}|n]}{[1\,n-1][n-1\,n][n1]}
\end{aligned}
\end{equation}
using momentum conservation in the second step. Combining this with $F$ in equation~(\ref{Fdef}) gives a factor
\begin{equation}
-\frac{\prod_{k=2}^{n-2}\langle k|p_{k+1}+p_{k+2}+\cdots +p_{n-1}|n]}{[1\,n-1][n-1\,n][n1]}
\end{equation}
Next, by carefully altering the limits of the products, we may re-express $N(n)$ as
\begin{equation}
		N(n) = \prod_{i=1}^{n-1}\prod_{j=i+1}^n\ [ij] 
		=-C(n) \left\{\prod_{i=1}^{n-3}\prod_{j=i+2}^{n-1}\ [ij] \right\}\prod_{k=2}^{n-2}\ [kn]
\label{Nmanipulate}
\end{equation}
where $C(n)$ is the cyclic product $[12][23]\cdots[n-1\,n][n1]$. The term in braces now cancels an identical term in the
numerator of~(\ref{BGK}). Hence we obtain
\begin{multline}
		\cM_{\rm BGK}(1^+,2^-,3^-,\ldots, n-1^-,n^+)=\frac{\kappa^2}{\hbar}\,\delta\left(\sum p\right)\\
		\times\ \left\{\frac{[1n]^8}{[1\,n-1][n-1\,n][n\,1]}\frac{1}{C(n)}\prod_{k=2}^{n-2}
		 \frac{\langle k|p_{k+1}+\cdots +p_{n-1}|n]}{[kn]}+ {\rm P}_{(2,\ldots,n-2)}\right\}\ ,
\end{multline}
which is the form of the amplitudes in equation~(\ref{amplitudes}).


\section{Yang-Mills}
\label{sec:ym}

In this appendix, we will review the twistor construction of the
Parke-Taylor amplitudes in Yang-Mills
theory (see~\cite{Mason,BMS} for further details). Although this section is not strictly
necessary for an understanding of the gravitational case, there are
nonetheless many analogies between the two and some readers may find
it useful to refer here for comparison. 

\subsection{Scattering off an Anti Self-Dual Yang-Mills Background}
\label{sec:ymgen}

On spacetime, Yang-Mills theory may be described by the Chalmers \&
Siegel~\cite{ChS} action
\begin{equation}
		S[A,G^+]=\frac{1}{g^2}\int_M {\rm tr} \left(G^+\wedge F - G^+\wedge G^+\right)\ ,
\label{ChS}
\end{equation}
where $F=dA+A^2$ and $G^+$ is a Lie algebra-valued self-dual 2-form. We will frequently drop the superscript from 
$G^+$, but it is always self-dual. The field equations are
\begin{equation}
		G^+=\frac{1}{2} F^+\qquad\hbox{and}\qquad
		D_AG^+=0\ ,
\label{ymeom}
\end{equation}
where $D_A$ is the covariant derivative. The first of these equations may be viewed as a constraint; enforcing it in~\eqref{ChS} one recovers the standard Yang-Mills action, upto a topological term. Using the Bianchi identity, the second equation is the standard Yang-Mills equations ${D_A}^*F=0$.

Anti self-dual solutions to~\eqref{ymeom} have $F^+=0$. Replacing $A\to A+a$ and $G\to G+{\rm g}$ and expanding the full field equations to linear order, one finds 
\begin{equation}
			2{\rm g}=(D_A a)^+\qquad\hbox{and}\qquad
			D_A{\rm g}=0
\label{ymlineom}
\end{equation}
when the background is anti self-dual. The solution space of these linear equations is an (infinite dimensional) vector space $U$ (to be considered modulo gauge transformations). If $\cR$ denotes the space of solutions to the full equations~\eqref{ymeom}, then $U$ may be interpreted as the fibre of $T\cR$ over a particular ASD solution. As for gravity, we identify $U^-\subset U$ as the subspace with ${\rm g}=0$. Since $F\to F+D_Aa = F+(D_Aa)^++(D_Aa)^-$, equation~\eqref{ymlineom} shows that linearized solutions in $U^-$ preserve the anti self-duality of the Yang-Mills curvature. $U^+$ is defined asymmetrically to be $U^+:=\{{\rm g}\in\Omega^{2+}(M,{\rm End} E)\,\big|\, D_A{\rm g}=0\}$, modulo gauge transformations. From equation~\eqref{ymlineom}, such ${\rm g}$ fields generate linear fluctuations in the self-dual part of the curvature. However, on an ASD background it does not make sense to ask for $U^+$ to be the solutions that are {\it purely} self-dual, because under a background gauge transformation with parameter $\chi$, the variation 
$a\to a+D_A\chi$ implies
\begin{equation}
		D_Aa\to D_Aa+D_A(D_A\chi)= D_Aa+[F^-,\chi]
\end{equation}
so that requiring $(D_Aa)^-=0$ would not be invariant under background gauge transformations. Again, this is summarized by the exact sequence
\begin{equation}
		0\to U^-\to U \to U^+\to 0
\label{seq}
\end{equation}
where $U^-\to U$ is an inclusion and the map $U\to U^+$ is $(a,{\rm g})\mapsto {\rm g}$. The fact that a self-dual fluctuation may or may not have an anti self-dual component obstructs the global splitting $U=U^-\oplus U^+$. Once again, this obstruction may be attributed to the MHV amplitudes, interpreted as scattering a linearized self-dual field off the ASD background.

Evaluating the action~\eqref{ChS} on $(A,G^+)=(A_0+a,{\rm g})$ where $(A_0,0)$ are an ASD background and $(a,{\rm g})$ obey the linearized equations~\eqref{ymlineom}, we find
\begin{equation}
		\frac{\im}{\hbar}S[A_0+a,{\rm g}]  = \frac{\im}{g^2\hbar}\int_M {\rm tr}\left({\rm g}\wedge{\rm g}\right)
\end{equation}
which, according to the path-integral argument in section~\ref{sec:gravity}, is the tree-level amplitude for a positive helicity  gluon to scatter off the background and emerge with negative helicity. We can again confirm this with a separate calculation.

The space of solutions $\cR$ of~\eqref{ymeom} again possesses a naturally defined closed two-form
\begin{equation}
		\Omega := \frac{1}{g^2}\int_C {\rm tr}\left( \delta G\wedge\delta A\right)\ .
\label{ymsymplectic}
\end{equation}
As a consequence of the field equations, $\Omega$ is independent of the Cauchy surface $C$ and descends to a symplectic form on $\cY/{\rm gauge}$. If $\Scr{A}_{1,2}=(a_{1,2},{\rm g}_{1,2})$ are two sets of linearized solutions, then
$g^2\Omega(\Scr{A}_1,\Scr{A}_2) = \int_C{\rm tr}\left( {\rm g}_1\wedge a_2-{\rm g}_2\wedge a_1\right)$. As for gravity on an anti self-dual spacetime, the symplectic form~\eqref{ymsymplectic} can be used to define a splitting of $U$ that depends on a choice of Cauchy surface $C$. Clearly, $U^-$ forms a Lagrangian subspace with respect to~\eqref{ymsymplectic} and we can ensure $U^+$ is likewise Lagrangian by {\it defining} a fluctuation $\Scr{A}_2$ to be purely self-dual if
\begin{equation}
		\Omega(\Scr{A}_2,\Scr{A}_1) = -\frac{1}{g^2}\int_C {\rm tr}\left({\rm g_2}\wedge a_1\right)
\end{equation}
for an {\it arbitrary} fluctuation $\Scr{A}_1$.

The quantum mechanical inner-product is defined in the same way as in the text (on Minkowski space $\bM$ the positive/negative frequency splitting can be performed straightforwardly) and agrees with the symplectic form~\eqref{ymsymplectic} on positive frequency states. The amplitude for a linearized fluctuation $\Scr{A}_1$ that has positive helicity and positive frequency at $\scri^-$ to emerge at $\scri^+$ with {\it negative} helicity (and positive energy) after traversing region of anti self-dual Yang-Mills curvature (in $\bM$) is $\langle\Scr{A}_2|\Scr{A}_1\rangle_{\rm asd}$, where $\Scr{A}_2$ is purely self-dual at $\scri^+$. In exact analogy to equation~\eqref{background}, we find
\begin{equation}
\begin{aligned}
		\big\langle\Scr{A}_2\big|\Scr{A}_1\big\rangle_{\rm asd} 
		&= \frac{\im}{g^2\hbar}\int_{\scri^+}{\rm tr}\left({\rm g}_2\wedge a_1\right)\\
		&= \frac{\im}{g^2\hbar}\int_{\bM}{\rm tr}\left(D_A {\rm g}_2 \wedge a_1 + {\rm g}_2\wedge D_A a_1\right)
		+\frac{\im}{g^2\hbar}\int_{\scri^-}{\rm tr}\left({\rm g}_2\wedge a_1\right)\\
		&=\frac{\im}{g^2\hbar}\int_{\bM}{\rm tr}\left({\rm g_2}\wedge {\rm g}_1\right)
\end{aligned}
\label{ymgen}
\end{equation}
after using the linearized field equations~\eqref{ymlineom} and the fact that $\Scr{A}_1$ is purely self-dual at $\scri^-$.

Equation~\eqref{ymgen} is a generating function for the Parke-Taylor amplitudes. To obtain them in their usual form, one must construct a background ASD field $A$ that is a (nonlinear) superposition of $n-2$ plane waves and solve the equation $D_A{\rm g}=0$ with such an $A$. Finally, one must  expand the above integral to the appropriate order. As for gravity, these problems are considerably simplified by the use of twistor theory, which brings out the integrability of the ASD Yang-Mills equations.


\subsection{The Twistor Theory of Yang-Mills}
\label{sec:twym}

For the basic notation of twistor space, we refer to the beginning of section~\ref{sec:twistors}. Anti self-dual connections on spacetime correspond to holomorphic
bundles $E$ on twistor space, by the Ward construction~\cite{Ward}. 
In the Dolbeault framework used in this
paper, such a bundle is determined by an operator $\delbar + \cA$
satisfying $\cF^{(0,2)}:=(\delbar+\cA)^2=0$, where $\delbar$ is the
standard $\delbar$-operator on twistor space and $\cA$ is the
$(0,1)$-form part of a connection on $E$ (and has homogeneity degree
0).  Note that $\delbar+\cA$ may be regarded as a deformation of the
$\delbar$-operator on a flat gauge bundle, while the integrability
condition $\cF^{(0,2)}=0$ arises as the field equations of the action
\begin{equation}
		\int_{\bP\bT^\prime}\Omega\wedge{\rm tr} (\cG\wedge\cF)
\label{BFact}
\end{equation}
where $\cG$ is a $(0,1)$-form of homogeneity $-4$ with values in ${\rm End}(E)$ and $\Omega$ is the canonical holomorphic (3,0)-form of
weight $+4$ on $\bP\bT^\prime$. Thus, (\ref{BFact}) is the twistor
equivalent of the $g^2\to0$ limit of~(\ref{ChS}) on spacetime.

Following Sparling~\cite{Sparling}, the spacetime Yang-Mills
connection can be reconstructed by first solving
\begin{equation}
		\left.\left(\delbar  + \cA\right)\right|_{L_x} H=0\, ,
\label{Sparling}
\end{equation}
where, for a Yang-Mills field on spacetime with gauge group $G$, $\cA$
takes values in the complexified Lie algebra of $G$ whereas $H$ is
valued in the complexification of $G$ itself.  The notation
$\left.(\delbar+\cA)\right|_{L_x}$ means the restriction of the
twistor space operator $\delbar+\cA$ to $L_x$.  A solution $H$
of~(\ref{Sparling}) is a global holomorphic frame of $E|_{L_x}$,
related to the twistor connection one-form by
\begin{equation}
		\cA|_{L_x}=-\delbar H\, H^{-1}\ .
\label{YMpuregauge}
\end{equation}
The generic existence of such frames for each $x$ is guaranteed by standard properties of holomorphic vector bundles\footnote{The Penrose-Ward transform requires $E|_{L_x}$ to be trivial. This will generically be the case and arises because the fibre of the Yang-Mills bundle over a spacetime point $x$ is by definition the space of global holomorphic sections of $E|_{L_x}$; these `jump' if 
$E|_{L_x}$ becomes non-trivial, so any twistor bundle that comes from a spacetime bundle will necessarily be trivial over $L_x$.}.  To reconstruct the spacetime connection $A$, first note that
$H^{-1}\pi^{\dot\alpha}\del H/\del x^{\alpha\dot\alpha}$ has homogeneity one in
$\pi_{\dot\alpha}$. Moreover, $H^{-1}\pi^{\dot\alpha}\del H/\del x^{\alpha\dot\alpha}$ is holomorphic on $L_x$, since
\begin{equation}
		\delbar\left(H^{-1}\pi^{\dot\alpha}\frac{\del H}{\del x^{\alpha\dot\alpha}}\right) =
		 H^{-1}\cA\,\pi^{\dot\alpha}\frac{\del H}{\del x^{\alpha\dot\alpha}}-H^{-1}\pi^{\dot\alpha}
		\frac{\del\ }{\del x^{\alpha\dot\alpha}}(\cA H)=0\ ,
\label{Hhol}
\end{equation}
where $\pi^{\dot\alpha}\del \cA/\del x_{\alpha\dot\alpha}=0$ because $\cA$ has been pulled back from $\bP\bT^\prime$ 
and so depends on $x$ only through the combination $x^{\alpha\dot\alpha}\pi_{\dot\alpha}$. Thus 
$H^{-1}\pi^{\dot\alpha}\del_{\alpha\dot\alpha}H$ must in fact be linear in $\pi_{\dot\alpha}$ and so may be written as  
\begin{equation}
		H^{-1}\pi^{\dot\alpha}\del_{\alpha\dot\alpha}H=\pi^{\dot\alpha}A_{\alpha\dot\alpha}(x)
\label{A-def}
\end{equation} 
for some Lie-algebra valued functions $A_{\alpha\dot\alpha}$ that depend only on spacetime. This provides the spacetime connection $A=A_{\alpha\dot\alpha} dx^{\alpha\dot\alpha}$.

\smallskip

To construct a twistor expression for $\langle\Scr{A}_2|\Scr{A}_1\rangle$, recall that for a flat Yang-Mills bundle, the Penrose transform of a linearized fluctuation ${\rm g}$ is related to $\cG$ by
\begin{equation}
		{\rm g}_{\dot\alpha\dot\beta}(x)=\int_{L_x}[\pi\,\rd\pi]\wedge\pi_{\dot\alpha}\pi_{\dot\beta}\, p^*\left(\cG\right)
\label{abelian}
\end{equation}
where ${\rm g}={\rm g}_{\dot\alpha\dot\beta}dx^{\alpha\dot\alpha}\wedge dx_{\alpha}^{\ \dot\beta}$. Moreover, if 
$\delbar\cG=0$ then ${\rm g}_{\dot\alpha\dot\beta}$
automatically obeys $\del^{\alpha\dot\alpha}{\rm g}_{\dot\alpha\dot\beta}=0$, again because the pullback 
$p^*\cG$ depends on $x$ only through $x^{\alpha\dot\alpha}\pi_{\dot\alpha}$. The equations 
$\delbar\cG=0$ and $\del^{\alpha\dot\alpha}G_{\dot\alpha\dot\beta}=0$ are the linearized field equations of~(\ref{BFact}) and~(\ref{ChS}) in the case that the background bundles are flat\footnote{They can also be thought of as Abelianized versions of the full theory.} so that we can find a gauge where $\cA=0$ and $A=0$. However, on a ASD Yang-Mills background, (\ref{abelian}) does not quite make sense. In order to add up an ${\rm End}(E)$-valued form over $L_x$, in the presence of a non-flat Yang-Mills bundle we need first to pick a holomorphic trivialization of $E|_{L_x}$ that is global over $L_x$: this is just the solution $H$ of equation~(\ref{Sparling}). The background-coupled twistor integral formula for 
${\rm g}^+$ is then
\begin{equation}
		{\rm g}_{\dot\alpha\dot\beta}(x)
		=\int[\pi\,{\rm d}\pi]\wedge \pi_{\dot\alpha}\pi_{\dot\beta}\,H^{-1}(x,\pi)\,p^*\left(\cG\right)\, H(x,\pi)\ .
\label{Gtrans}
\end{equation}
From equation~(\ref{Gtrans}) we now find (dropping the pullback symbol $p^*$)
\begin{eqnarray}
		\del^{\alpha\dot\alpha}{\rm g}_{\dot\alpha\dot\beta}&=&\int[\pi\,\rd\pi]\wedge \pi_{\dot\alpha}\pi_{\dot\beta}
		\frac{\del\ }{\del x_{\alpha\dot\alpha}}\left(\,H^{-1}\cG H\right)\\
			&=&\int[\pi\,\rd\pi]\wedge \pi_{\dot\alpha}\pi_{\dot\beta}
			\left(-\pi_{\dot\alpha}A^{\alpha\dot\alpha}H^{-1}\cG H+H^{-1}\cG H A^{\alpha\dot\alpha}\right)
			=-\left[A^{\alpha\dot\alpha},{\rm g}_{\dot\alpha\dot\beta}\right] \nonumber
\end{eqnarray}
or in other words $D_A{\rm g}=0$, which is the linearized field
equation~\eqref{ymlineom} for ${\rm g}$ on an ASD background.
Therefore, the scattering amplitude we seek is given by
\begin{equation}
		\big\langle\Scr{A}_2\big|\Scr{A}_1\big\rangle_{\rm asd}
		=\frac{\im}{g^2\hbar}\int{\rm d}^4x\ [\pi_1\,{\rm d}\pi_1][\pi_2\,{\rm d}\pi_2]\ [\pi_1\,\pi_2]^2\,
				{\rm tr}\left(H^{-1}_2\cG^{\phantom{2}}_2 H_2^{\phantom{2}}\,
					H^{-1}_1\cG_1^{\phantom{1}} H_1^{\phantom{1}}\right)  
\label{twistorYM}
\end{equation}
where the integral on the right is then taken over $\bR^4\times\bC\bP^1\times\bC\bP^1$.

To obtain the Parke-Taylor amplitudes we must expand the frames $H$ as a perturbation series around a flat background 
by inverting the relation $\cA|_{L_x}=-\delbar H H^{-1}$. Rather than do this directly (see~\cite{SelivanovYM}), it is simpler to note that the Green's function $K_{12}$ for the $\delbar$-operator on $L_x$, acting on sections of ${\rm End}(E)|_{L_x}$, is related to $H$ by
\begin{equation}
		K_{12}(x,\pi_1,\pi_2)= \left(\frac{1}{2\pi\im}\right)\frac{H(x,\pi_1)H^{-1}(x,\pi_2)}{[\pi_1\,\pi_2]}
\end{equation}
and may formally be thought of as $(\delbar+\cA|_{L_x})^{-1}$. This is analogous to equation~\eqref{delbary} in section~\ref{sec:twistamps}. The Green's function thus depends non-polynomially on $\cA$; expanding the right hand side of~(\ref{twistorYM}) as a series in $\cA$ using $\left.K_{ij}\right|_{\cA=0}=1/2\pi\im[\pi_i\,\pi_j]$, one obtains
\begin{equation}
		\frac{\im}{g^2\hbar}\int{\rm d}^4x\prod_{i=1}^n\frac{[\pi_i\,{\rm d}\pi_i]}{[\pi_i\,\pi_{i+1}]}\ 
		\sum_{p=2}^n[\pi_1\,\pi_p]^4\ {\rm tr}\left(\cA_n\cdots\cA_{p+1}\,\cG_p\,\cA_{p-1}\cdots\cA_{2}\,\cG_{1}\right)
\label{YMexpand} 
\end{equation}
for the vertex involving $n$ fields. To obtain the Parke-Taylor amplitudes, take $\cA$ and $\cG$ to be linear combinations of momentum eigenstates momentum $p_{\alpha\dot\alpha}=\tilde k_\alpha k_{\dot\alpha}$, with helicities $-1$ and $+1$, respectively. As in~\cite{CSW}, these can be represented by the twistor functions,
\begin{equation}
\begin{aligned}
		\cA&=g\sum_{i=3}^{n} \epsilon_i T_i\ \bar\delta_{(0)}([\pi\, i]) 
		\exp\left(\frac{\langle \omega\, i\rangle[i\,\sigma]}{[\pi\,\sigma]}\right)\\ 
		\cG&=g\sum_{i=1}^{2} \epsilon_i T_i\ \bar \delta_{(-4)}([\pi\, i])
		\exp\left(\frac{\langle\omega\, i\rangle[i\,\sigma]}{[\pi\,\sigma]}\right)\, ,
\end{aligned}
\label{AGmom}
\end{equation}
where $T_i$ are arbitrary elements of the Lie algebra of the gauge group, and the $\epsilon_i$ are expansion
parameters. The coefficient of $\prod_{i=1}^n\epsilon_i$ in~(\ref{YMexpand}) is the $n^{\rm th}$-order Parke-Taylor amplitude (complete with the appropriate colour-trace).

Treating the fields $\cA$ and $\cG$ as ${\rm End}\,E$-valued (0,1)-forms, rather than representatives of cohomology classes, we can combine~\eqref{YMexpand} with the action~\eqref{BFact} to obtain a twistor action for the MHV diagram formulation of Yang-Mills. It is straightforward to extend this to an action for $\cN=4$ SYM. See~\cite{Mason,BMS,BMS2} for details.

\end{document}